\documentclass[aps,onecolumn,11pt,floatfix,altaffilletter,superscriptaddress,
               preprintnumbers,tightenlines,showpacs,showkeys,notitlepage,nofootinbib,preprint]{revtex4-1}

\usepackage{amsfonts, amsmath, amssymb, graphicx, color, hyperref}
\usepackage{multirow}
\usepackage{cleveref}
\usepackage{physics}
\usepackage{tikz}
\usetikzlibrary{shapes.geometric, positioning, fit, backgrounds}
\usepackage{outlines}
\usepackage{enumitem}
\setlist{nosep}

\usepackage{mmacells}
\mmaDefineMathReplacement[≤]{<=}{\leq}
\mmaDefineMathReplacement[≥]{>=}{\geq}
\mmaDefineMathReplacement[≠]{!=}{\neq}
\mmaDefineMathReplacement[⧴]{:>}{:\hspace{-.2em}\to}[2]
\mmaDefineMathReplacement{∉}{\notin}
\mmaDefineMathReplacement{∞}{\infty}
\mmaDefineMathReplacement{`}{\grave{}}
\mmaDefineMathReplacement{ϕ}{\phi}
\mmaDefineMathReplacement{γ}{\gamma}
\mmaDefineMathReplacement{λ}{\lambda}
\mmaDefineMathReplacement{κ}{\kappa}
\makeatletter
\mmaSet{
    morefv={formatcom*={%
      \refstepcounter{lstlisting}%
      \ifx\lst@label\@empty\else
        \label{\lst@label}
      \fi
      \ifx\lst@@caption\@empty\else
        \addcontentsline{lol}{lstlisting}{\protect\numberline{\thelstlisting}\lst@@caption}
      \fi
    }},
    linklocaluri=mma/symbol/definition:#1,
    leftmargin=3em,
    labelsep=.5em,
    }
\makeatother

\usepackage[utf8]{inputenc}

\definecolor{nicered}{rgb}{.7,.1,.1}
\definecolor{nicegreen}{rgb}{.1,.5,.1}
\definecolor{darkblue}{rgb}{0,0,.5}
\hypersetup{colorlinks, citecolor=nicegreen, linkcolor=darkblue, urlcolor=darkblue}

\newcommand{\packageName}{\texttt{PT2GWFinder}}

\newcommand{\di}{\text{d}}
\newcommand{\FV}{\text{FV}}
\newcommand{\TV}{\text{TV}}
\newcommand{\epsal}{\epsilon_\alpha}
\newcommand{\Math}{\texttt{Mathematica}}

\makeatletter
\renewcommand\frontmatter@abstractwidth{\dimexpr\textwidth\relax}
\makeatother

\begin{document}

\title{%
\texorpdfstring{\includegraphics[width=0.3\linewidth]{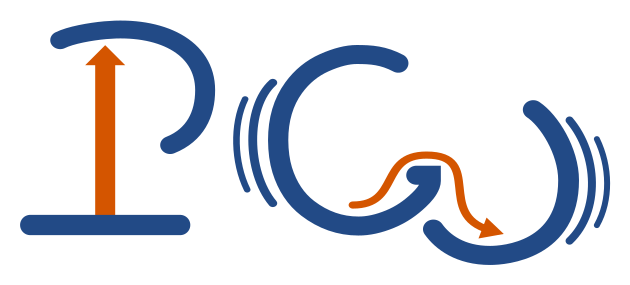}\\[1cm]}{}
\texttt{PT2GWFinder}: A Package for Cosmological First-Order Phase Transitions and Gravitational Waves}

\author{Vedran Brdar}
\email{vedran.brdar@okstate.edu}
\affiliation{Department of Physics, Oklahoma State University, Stillwater, OK, 74078, USA}

\author{Marco Finetti}
\email{mfinetti@ua.pt}
\affiliation{Departamento de Física, Universidade de Aveiro and CIDMA, Campus de Santiago, 3810-183 Aveiro, Portugal}
\affiliation{Laboratório de Instrumentação e Física Experimental de Partículas (LIP), Universidade do Minho, 4710-057 Braga, Portugal}
\affiliation{
    Department of Mathematics and Physics, University of Stavanger,
    NO-4036 Stavanger,
    Norway
    }

\author{Marco Matteini}
\email{marco.matteini@ijs.si}
\affiliation{Jo\v{z}ef Stefan Institute, Jamova 39, 1000 Ljubljana, Slovenia}
\affiliation{Faculty of Mathematics and Physics, University of Ljubljana, Jadranska 19, 1000 Ljubljana, Slovenia}

\author{António P. Morais}
\email{amorais@fisica.uminho.pt}
\affiliation{Departamento de Física, Escola de Ciências, Universidade do Minho, 4710-057 Braga, Portugal}
\affiliation{Laboratório de Instrumentação e Física Experimental de Partículas (LIP), Universidade do Minho, 4710-057 Braga, Portugal}

\author{Miha Nemev\v{s}ek}
\email{miha.nemevsek@ijs.si}
\affiliation{Jo\v{z}ef Stefan Institute, Jamova 39, 1000 Ljubljana, Slovenia}
\affiliation{Faculty of Mathematics and Physics, University of Ljubljana, Jadranska 19, 1000 Ljubljana, Slovenia}


\begin{abstract}
The detection of gravitational waves from binary black hole and neutron star mergers by ground-based interferometers, as well as the evidence for a gravitational wave background from pulsar timing array experiments, has marked a new era in astrophysics and cosmology. These experiments also have great potential for discovering new physics through gravitational wave detection. One of the most motivated sources of gravitational waves that can be realized only within a beyond-the-Standard-Model framework is first-order phase transitions. In this work we release \texttt{PT2GWFinder}, a \texttt{Mathematica} package designed to compute phase transition parameters and the gravitational wave power spectrum for an \textit{arbitrary} scalar theory exhibiting a first-order phase transition, in
scenarios where a single scalar acquires a vacuum expectation value. \texttt{PT2GWFinder} performs the phase tracing, computes the bounce profile and action using \texttt{FindBounce}, calculates the relevant temperatures and phase transition parameters, and finally evaluates the gravitational wave spectrum. Additionally, it offers a user-friendly interface with \texttt{DRalgo}, which enables the computation of the dimensionally reduced effective potential in the high-temperature regime. This work includes a user manual and two models that demonstrate the capability and performance of \texttt{PT2GWFinder}. As a supplement, for one of these models we obtain the bounce solution and action analytically in the thin-wall approximation and demonstrate excellent agreement with the numerical approach.

\vspace{5mm}
\noindent\textbf{Program summary}\\
\textit{Program title}: \texttt{PT2GWFinder}\\
\textit{Developer's repository link}: \url{https://github.com/finshky/PT2GW}\\
\textit{Licensing provisions}: GNU General Public License 3\\
\textit{Programming language}: \Math\\
\textit{Nature of problem}: Search and characterization of cosmological first-order phase transitions, computation of the generated gravitational wave spectra.\\
\textit{Solution method}: Construction of the Euclidean bounce action function, using \texttt{FindBounce}, and application of integral criteria to determine the transition temperatures.\\
\textit{Restrictions}: Mathematica version 13 or above, applicable to single-field models.
\end{abstract}

\maketitle
\tableofcontents

\section{Introduction} \label{sec:intro}
\noindent
The discovery of gravitational waves (GWs) has opened a new avenue for exploring the history and evolution of the Universe.
One example of a process that generates gravitational radiation is a cosmological first-order phase transition (FOPT) \cite{Witten:1984rs}, where the Universe decays from a metastable configuration (false vacuum) to a stable one (true vacuum).
A FOPT proceeds via bubble nucleation.
Thereby, spherical regions of spacetime decay to the stable phase, creating an interface (the bubble wall) that expands under the pressure exerted by the difference in potential energy between the phases and experiences friction due to the interaction with the cosmic fluid.
As the bubbles expand, they collide with each other and convert the entire spacetime into the true vacuum configuration.
The fluid is set in motion by the expanding bubbles, in the form of sound waves and, possibly, magnetohydrodynamic (MHD) turbulence.
Together with the colliding bubble walls, these generate an anisotropic stress that acts as a source of gravitational radiation.
The signal today would appear as a stochastic gravitational wave background (SGWB).
Cosmological FOPTs are directly linked to beyond-the-Standard-Model (BSM) physics, as, within the Standard Model, the electroweak (EW) phase transition and the QCD phase transition are crossovers~\cite{Kajantie:1995kf, Stephanov:2006zvm}.
A number of current and future gravitational wave experiments, such as 
LIGO~\cite{LIGOScientific:2014pky}, VIRGO~\cite{VIRGO:2014yos}, KAGRA~\cite{KAGRA:2020tym}, CE~\cite{Reitze:2019iox}, SKA~\cite{2013CQGra..30v4011L}, 
IPTA~\cite{Hobbs:2009yy},
NANOGrav~\cite{McLaughlin:2013ira}, EPTA~\cite{Kramer:2013kea}, PPTA~\cite{manchester2013parkes}, 
LISA~\cite{LISA:2017pwj}, DECIGO~\cite{Kawamura_2011}, BBO~\cite{Corbin:2005ny}, and ET~\cite{Punturo:2010zz} 
might be able to detect gravitational waves coming from FOPTs, depending on the typical frequency and amplitude of the signal.
A robust control of the dynamics of the phase transition is key to producing reliable predictions for the potentially observable signals.

In light of this, we release the \Math\ package \packageName. For \textit{any} given scalar particle physics model, where the phase transition proceeds along 
a single scalar direction, the package performs the phase tracing and calculates the bounce profile and the associated bounce action via \texttt{FindBounce}.
It determines the critical, nucleation, and percolation temperatures, as well as the phase transition strength and duration. 
It calculates the GW spectrum from bubble collisions, sound waves and MHD turbulence using the latest available templates~\cite{Caprini:2024hue}.
The inputs required for the calculation are the effective potential and the velocity at which the bubble walls expand.
The efficiency factor for the production of GWs, coming from bubble collisions, is set to 0 by default, while the calculation of the efficiency from sound waves follows Ref.~\cite{Espinosa:2010hh}.
As an alternative, the user can provide a numerical value for the bubble walls efficiency or a list and properties of the particles that acquire a mass by crossing the bubble wall. 
The computation of the efficiency factors then follows Refs.~\cite{Ellis:2019oqb,Ellis:2020awk}.

Currently, three other codes that share common goals are available: \texttt{BSMPTv3} \cite{Basler:2024aaf} and \texttt{PhaseTracer2}~\cite{Athron:2024xrh} written 
in~\texttt{C++}, and the Python package \texttt{ELENA}~\cite{Costa:2025pew}.
All of them come with preloaded models, and allow for user-defined model implementation.
In \texttt{BSMPTv3} the user must provide the particle content, potential parameters, counterterms and vacuum expectation values (VEVs), and the code calculates the effective thermal potential internally using daisy resummation~\cite{Parwani:1991gq, Arnold:1992rz}.
Since version 3 (BSMPTv3), the phase tracing and bounce computation follow a similar approach to 
\texttt{CosmoTransitions}~\cite{Wainwright:2011kj}, with improved stability and speed.
In \texttt{PhaseTracer2}, the treatment of thermal corrections can be done either with daisy resummation or dimensional reduction~\cite{GINSPARG1980388,PhysRevD.23.2305}.
The phase tracing is similar to \texttt{CosmoTransitions} but faster.
For the bounce calculation, different algorithms are available, including an interface with \texttt{BubbleProfiler}~\cite{Athron:2019nbd}.
At its present stage, \texttt{PhaseTracer2} determines the GW spectrum at the nucleation temperature using a simple heuristic estimate.

The companion package \texttt{TransitionSolver} provides a more rigorous treatment of the transition dynamics.
In their latest versions, both \texttt{BSMPTv3} and \texttt{PhaseTracer2} implement the computations of phase transition and GW parameters.

The recently introduced \texttt{ELENA} implements the tunneling potential formalism---with a vectorized implementation---rather than the traditional bounce solver.
As for \packageName, the first public release targets single-field potentials. The finite-temperature scalar potential is built from the \texttt{generic\_potential} class of \texttt{CosmoTransitions}, allowing already-implemented models to be used directly.

Our package \packageName\ provides a comprehensive pipeline from a particle physics model to the resulting gravitational wave spectrum, while remaining agnostic to the specific method used for constructing the effective potential.
Any scalar potential that exhibits a FOPT,
where a single field is allowed to acquire a VEV
is suitable\footnote{
    Other scalars can be present in the theory, but only one can acquire a VEV.
    If by performing a global rotation, the problem can be reduced to that of a single scalar acquiring a VEV, this procedure should be performed in the construction of the thermal potential. \packageName\ can handle multiple-step PTs.
}. The user is required to supply the full thermal potential explicitly.
In particular, potentials derived through dimensional reduction are fully supported.
For this purpose, we have implemented the helper file \texttt{DRTools}, which allows us to 
retrieve useful information from the output of \texttt{DRalgo}~\cite{Ekstedt:2022bff} 
and plug it directly into our package.
The phase tracing offers different possibilities, from a semi-analytical derivation of the 
phases as a function of temperature, to a numerical minimization of the potential at a 
given set of temperatures, depending on the simplicity of the potential as a function 
of the field and temperature.
The calculation of the bounce profile and action utilizes the robust 
\texttt{FindBounce}~\cite{Guada:2020xnz} package, which is based on the polygonal 
bounce method~\cite{Guada:2018jek}.
The bubble wall velocity must be provided as an external input. For recent developments in the computation of this quantity, we refer to Refs.~\cite{Konstandin:2014zta,Kozaczuk:2015owa,DeCurtis:2022hlx,Jiang:2022btc,DeCurtis:2024hvh}. For its numerical derivation in non-runaway scenarios, we direct the reader to \texttt{WallGo} \cite{Ekstedt:2024fyq}.
The user can implement custom templates for the efficiency factors and the GW power 
spectra by manually modifying the \texttt{GW} module of the package.
If the particle physics model is simple enough, such that the potential is a polynomial 
in the scalar field up to the fourth power, a robust analytical control over the 
Euclidean action is provided by the thin-wall approximation, even away from the strict 
thin-wall regime.
For more details, we refer the reader to Refs.~\cite{Ivanov:2022osf, Matteini:2024xvg} and 
to the~\ref{app:analytics}.

This paper is organized as follows.
In Section~\ref{sec:PTs_general} we introduce the physics of cosmological phase transitions, together with the relevant quantities that describe such processes. 
In Section~\ref{sec:download_install_example}, we provide details on downloading and installing the package, along with a simple first example of its usage.
In Section~\ref{sec:method_description} we outline the pipeline of \packageName\ and 
describe the numerical methods employed.
In Section~\ref{sec:examples} we apply \packageName\ to study phase transitions for two
phenomenologically relevant models: the coupled fluid-scalar field (CFF) model and the Dark Abelian Higgs model, in order to demonstrate successful usage in realistic scenarios.
In Section~\ref{sec:usage_manual}, we provide a user's manual for \packageName\, including a detailed description of all the functions and the available options.
We summarize and provide an outlook for future work in Section~\ref{sec:conclusion}.

\section{General features of a phase transition}\label{sec:PTs_general}
\noindent
The first step in analyzing a cosmological phase transition for a given model is 
to calculate the effective potential.
In a finite temperature field theory, the perturbative expansion is expected to break down 
due to infrared divergences.
In a scalar theory, this is due to diagrams where the IR modes are screened by the UV modes,
known as daisy diagrams, which dominate at each loop order, causing the self-coupling to 
diverge at high temperatures.
There have typically been two methods to solve this problem and obtain the thermal potential 
for a given field theory: daisy resummation and dimensional reduction.
Daisy resummation consists of resumming the diverging diagrams, together with introducing 
a mass counterterm that depends on temperature.
Two prescriptions exist to implement such resummation. 
The Parwani procedure~\cite{Parwani:1991gq} replaces the mass terms for scalars and 
longitudinal gauge bosons, appearing in the effective potential, with their resummed version 
$m^2_i(\phi)\to m^2_i(\phi)+\Pi_i(T)$, where $\Pi_i(T)$ is the Debye mass. 
The Arnold-Espinosa procedure~\cite{Arnold:1992rz} introduces a daisy correction term 
in the potential of the form
\begin{align}
  V_\text{daisy}(S,T)=-\frac{T}{12\pi}\sum_i n_i \left[\left(m^2(\phi)+\Pi(T)\right)_i^{3/2}-\left(m^2(\phi)\right)_i^{3/2}\right] \, ,
\end{align}
where $n_i$ is the number of degrees of freedom for each particle. 
The first term implements the resummation, while the second removes the double counting.
An alternative method to cure the IR divergences is the use of dimensional 
reduction~\cite{GINSPARG1980388,PhysRevD.23.2305}. 
Here, the UV modes (non-zero Matsubara modes in the imaginary time formalism) are integrated
out, leaving an effective field theory (EFT) for the IR modes that live in three spatial dimensions.
This technique can also be interpreted as a systematic way of performing thermal resummations
within perturbation theory~\cite{Farakos:1994kx}, where the leading order is equivalent to 
the resummation via thermal masses.

As the Universe cools down, the phase structure might change, typically in the form of the
appearance of a new phase that, at a certain temperature, becomes the stable vacuum.
The critical temperature $T_c$ is defined as the temperature at which two phases are degenerate.
Below this temperature, the new phase becomes the global minimum of the theory, and the possibility of a transition opens up.
If a potential barrier is present between the phases, the transition will be first-order and will proceed via bubble nucleation.

Once the minima of the potential are known, the next step is to calculate the solution of 
the equations of motion (known as the bounce, which is the spherically symmetric field configuration describing the tunneling) and the corresponding Euclidean action $S$.
For a transition at finite temperature, the time component of the 4-dimensional Euclidean 
action is given by $T^{-1}$, and the decay rate per unit volume then depends on the 
3-dimensional action $S_3$ via~\cite{Linde:1981zj}
\begin{align}
    \Gamma(T)\simeq T^4\left(\frac{S_3}{2\pi T}\right)^{3/2} \text{e}^{-S_3/T} \, ,
\end{align}
where the contribution from the fluctuations around the bounce is estimated by dimensional arguments to be $\sim T^3$.

In a radiation-dominated Universe, taking into account the vacuum energy density contribution, the Hubble parameter is given by~\cite{Espinosa:2010hh}
\begin{align} \label{eq:Hubble}
  H^2(T) = \frac{\rho_\gamma(T)+\rho_\text{vac}(T)}{3M_\text{PL}^2} 
  = \frac{1}{3M_\text{PL}^2}\left(\frac{\pi^2}{30}g_*T^4+\Delta V(T) \right) \, ,
\end{align}
where $M_\text{PL}$ denotes the reduced Planck mass, $M_\text{PL} = 2.435 \times 10^{18}$ GeV, and $g_*$ denotes the effective number of relativistic degrees of freedom in the thermal plasma.
The contribution from the vacuum energy density is given by the difference in potential between the two minima $\Delta V(T)~\equiv~V(\phi_\FV,T)-V(\phi_\TV,T)$.

The nucleation temperature $T_n$ is an indicator of the onset of the phase transition and is defined as the temperature at which, on average, one bubble of true vacuum nucleates per Hubble volume\footnote{
The integrand in Eq.~\eqref{eq:Tn_definition} assumes the false vacuum fraction is 1. For fast transitions, this is a reasonable approximation down to the nucleation temperature: $P_\mathrm{FV}(t_n<t<t_c)\approx1$.
}
\begin{align} \label{eq:Tn_definition}
    \int_{t_c}^{t_n}\di t \,\frac{\Gamma(t)}{H(t)^3}=\int_{T_n}^{T_c}\frac{\di T}{T} \frac{\Gamma(T)}{H(T)^4} = 1 \, .
\end{align}
Moving from time variable to temperature variable requires some assumptions \cite{Athron:2022mmm}. 
Usually, an adiabatic expansion of the Universe is assumed, i.e.
\begin{align}
  \dfrac{\di}{\di t}\left(s(t)a^3(t)\right) = 0 \implies \dfrac{\di s}{\di t} = -3H(t)s(t)\, ,
\end{align}
where $s$ is the entropy density of the plasma and $a$ is the scale factor of the Universe. Given
\begin{align}
    s(T) = \dfrac{\partial p}{\partial T} = -\dfrac{\partial V}{\partial T} \, ,
\end{align}
the adiabatic expansion condition can be written as
\begin{align}
    \dfrac{\di T}{\di t} = -3H(T)\dfrac{\partial_T V}{\partial^2_{T} V} \, .
\end{align}
From energy conservation, it is typical to assume a homogeneous energy density even when bubbles are present, and the derivatives are evaluated at the false vacuum.
A further assumption regarding the equation of state is required to proceed. A common one is the MIT bag equation of state \cite{Chodos:1974je}
\begin{align}
    V(T) = c_1T^4+c_2 \, ,
\end{align}
where $c_{1,2}$ are independent of temperature. Lower powers of the temperature are neglected in the high-temperature approximation that is used in this equation of state.

Alternatively, one can consider the simplified argument that the entropy density as a function of temperature is given by
\begin{align}
    s(T)\propto g_*(T)\, T^3 \, ,
\end{align}
where any non-trivial dependence on temperature from degrees of freedom that are not relativistic is included in the $g_*$.
This way avoids assumptions on the temperature dependence of the potential.

Then, the condition of adiabatic expansion of the Universe becomes
\begin{align} \label{eq:adiabatic_time_temp}
    \frac{\di T}{\di t}=-TH(T) \, ,
\end{align}
which gives the relation between time and temperature and justifies the first equivalence in \cref{eq:Tn_definition}.

An approximate criterion to obtain the nucleation temperature is that the number of bubbles in a Hubble volume $H^{-3}$ nucleated in a Hubble time $H^{-1}$ is unity
\begin{align} \label{eq:GammaH-4}
    \Gamma H^{-4}|_{T_n}\approx 1 \, .
\end{align}
The above can be rewritten in terms of the Euclidean action as
\begin{align} \label{eq:ApprCriterionTn}
    \frac{S_3}{T_n} \approx 4\log\left(\frac{T_n}{H(T_n)}\right) \, ,
\end{align}
which, for transitions at the EW scale and in radiation domination, is~\cite{Caprini:2019egz}
\begin{align} \label{eq:Tn_condition_EW}
    \frac{S_3}{T_n} \approx 140 \, . 
\end{align}

In order to understand whether the transition completes or not, we introduce the percolation temperature, defined as the temperature at which there is one connected region of true vacuum given by the merger of the bubbles.
This corresponds to the situation in which at least $34\%$ of the comoving volume has been converted to the true vacuum \cite{LIN2018299,LI2020112815}, or, equivalently, the probability of finding a point still in the false vacuum is 
\begin{equation} \label{eq:Tp_definition}
    P_\text{FV}\approx0.71\ .
\end{equation}
This probability is given by the JMAK equation\footnote{The probability $P_\text{FV}$ is equivalently the fractional volume of false vacuum. The exponent $I$ is sometimes referred to as the ``extended'' fractional volume of true vacuum \cite{Kolmogorov:1937,Johnson:1939,Avrami:1939}, for which the mutual overlap of bubbles is neglected.}
\cite{Kolmogorov:1937,Johnson:1939,Avrami:1939, Guth:1981uk, Ellis:2018mja}.
\begin{align}
  P_\text{FV}(t) &= e^{-I(t)} \, , & 
  I(t) &= \frac{4\pi}{3}\int_{t_c}^t \di t' \,\Gamma(t')\,a(t')^3\, r(t,t')^3 \, ,
\end{align}
where $t_c$ is the time corresponding to the critical temperature, $\Gamma$ is the nucleation rate per unit time per unit comoving volume, $a$ is the scale factor and
\begin{align}
    r(t,t') = \int_{t'}^t \di t'' \frac{v_w}{a(t'')}
\end{align}
is the comoving size of a bubble that nucleated at time $t'$ and has evolved until time $t$, with $v_w$ the bubble wall velocity, assumed to be constant.
These quantities can be expressed in terms of temperature, assuming an adiabatic expansion of the Universe. We start by noticing that \cite{Athron:2023xlk}
\begin{align}
    \frac{1}{a}\frac{\di a}{\di t}=H \implies \log\frac{a_2}{a_1}=\int_{t_1}^{t_2}\di t \, H(t)=\int^{T_1}_{T_2}\frac{\di T}{T}=\log\frac{T_1}{T_2} \implies a(T)\propto \frac{1}{T} \, ,
\end{align}
which leads to
\begin{align}
    a(T')r(T,T') = v_w a(T')\int^{T'}_T \frac{\di T''}{T''H(T'')a(T'')}=\frac{v_w}{T'}\int^{T'}_T \frac{\di T''}{H(T'')} \, ,
\end{align}
so that
\begin{align} \label{eq:I_percolation}
    I(T) = \frac{4\pi}{3} v_w^3\int^{T_c}_T \di T' \,\frac{\Gamma(T')}{H(T')\,T'^4}\left(\int_T^{T'}\frac{\di T''}{H(T'')}\right)^3 \, .
\end{align}

In order for the transition to complete, a stronger requirement is that, at the percolation temperature, the physical volume of false vacuum, $\mathcal{V}_\text{FV}=a^3\text{e}^{-I}$, is decreasing \cite{Ellis:2018mja,Athron:2022mmm}.
This might be relevant in scenarios where the Universe is vacuum-dominated and therefore experiencing an inflated expansion.
The condition for a decreasing false vacuum volume reads
\begin{align} \label{eq:perc_condition}
    \frac{1}{\mathcal{V}_\text{FV}}\frac{\di \mathcal{V}_\text{FV}}{\di t} = 3H(t)-\frac{\di I(t)}{\di t} = H(T)\left(3+T\,\frac{\di I(T)}{\di T}\right)<0 \, .
\end{align}

Eqs.~\labelcref{eq:I_percolation,eq:perc_condition} hold for an adiabatic expansion of the Universe, where reheating can be ignored. This is the case for supersonic detonations, while for subsonic deflagrations and hybrid expansion modes, Eq.~\eqref{eq:adiabatic_time_temp} will be modified \cite{Athron:2022mmm}. Furthermore, there are scenarios in which the condition \cref{eq:perc_condition} is not satisfied at percolation, but only at a lower temperature \cite{Ellis:2018mja,Goncalves:2024lrk}. Nonetheless, \cref{eq:perc_condition} can still impose strong constraints on the possible amount of supercooling.

\subsection{Phase transition parameters}
The phase transition strength is defined as the energy released by the vacuum transition, normalized to the radiation energy density, evaluated at the transition temperature $T_*$
\begin{align}\label{eq:alpha}
     \alpha = \frac{1}{\rho_\gamma(T_*)}\left[ V(\phi_\FV,T)-V(\phi_\TV,T)-\frac{T}{4}\left( \dfrac{\partial V}{\partial T}(\phi_\FV,T) - \dfrac{\partial V}{\partial T}(\phi_\TV,T) \right) \right]_{T=T_*} \, ,
\end{align}
where $\rho_\gamma(T)=\pi^2g_*T^4/30$ is the radiation energy density.

The second phase transition parameter is the inverse duration of the phase transition
\begin{align} \label{eq:beta}
    \beta/H_* =\left[ T \dfrac{\di}{\di T}\log\Gamma(T)\right]_{T=T_*}
\end{align}
evaluated at the temperature at which the transition ends $T_*$, which we generally take to be the percolation temperature.
In \packageName\ we use
\begin{equation}
   \beta/H_* \simeq \left[T \dfrac{\di}{\di T}\frac{S_3(T)}{T}\right]_{T=T_*} \, ,
\end{equation}
as the precise calculation of the subleading prefactor is beyond the scope of our paper.

Typical values for the phase transition strength range from $\alpha\sim \mathcal{O}(0.01)$ for weak transitions, to $\alpha\sim \mathcal{O}(0.1)$ for intermediate transitions and $\alpha \gtrsim \mathcal{O}(1)$ for strong transitions.
The duration of the phase transition is usually assumed to be less than a Hubble time, i.e. $\beta/H_* > 1$.
These two parameters are typically not completely independent. Stronger phase transitions, with larger values of $\alpha$, generally require more time to complete, leading to
smaller values of $\beta/H$.
On the contrary, weaker phase transitions usually proceed rather quickly.

 \subsection{Gravitational wave power spectrum} \label{subsec:GW_spectrum}
Gravitational waves are sourced by bubble wall collisions, sound waves in the cosmic fluid, and possibly MHD turbulence in the later stages of the phase transition.
Each contribution has a characteristic power spectrum, and the total power spectrum is given by the sum of the three contributions.
The amplitudes and frequencies depend on the temperature at the end of the transition $T_*$, the inverse duration $\beta/H_*$, the strength $\alpha$, and the bubble wall velocity $v_w$.
For the MHD turbulence contribution, the spectrum also depends on the energy fraction converted into turbulent motion $\varepsilon$.

Here, we consider GW templates from the literature \cite{Caprini:2024hue}, which model the contributions from bubble wall collisions with a broken power law, while those from sound waves and MHD turbulence are modelled by a double broken power law: 
\begin{equation} \label{eq:Omega}
\begin{split}
    h^2\Omega_\text{col}(f) &\sim \tilde{K}^2
        \left(\frac{H_*}{\beta}\right)^2
        \left(\frac{f}{f_p}\right)^{2.4}
        \left[\frac{1}{2}+\frac{1}{2}\left(\frac{f}{f_p}\right)^{1.2}\right]^{-4} \, , \\
    h^2\Omega_\text{sw}(f) &\sim K^2
        \left(\frac{H_*}{\beta}\right)
        \left(\frac{f}{f_1^{\text{sw}}}\right)^{3}
        \left[1+\left(\frac{f}{f_1^{\text{sw}}}\right)^{2}\right]^{-1}
        \left[1+\left(\frac{f}{f_2^{\text{sw}}}\right)^{4}\right]^{-1} \, ,\\
    h^2\Omega_\text{tur}(f) &\sim \varepsilon^2 K^2
        \left(\frac{H_*}{\beta}\right)^2
        \left(\frac{f_1^{\text{tur}}}{f_2^{\text{tur}}}\right)
        \left(\frac{f}{f_1^{\text{tur}}}\right)^{3}
        \left[1+\left(\frac{f}{f_1^{\text{tur}}}\right)^{4}\right]^{-1/2}
        \left[1+\left(\frac{f}{f_2^{\text{tur}}}\right)^{2.15}\right]^{-1.7} \, .
\end{split}
\end{equation}
For the broken power law, the spectrum grows like $f^{2.4}$ for $f\ll f_p$, and decreases as $f^{-2.4}$ for $f\gg f_p$, where $f_p$ is the peak frequency.
For the double broken power law, the spectrum behaves like $f^{n_1}$ for $f<f_1$, like $f^{n_2}$ for $f_1<f<f_2$, and like $f^{n_3}$ for $f>f_2$, with $n_1=3$, $n_2=1$, and $n_3=-3$ for sound waves and $n_3=-8/3$ for MHD turbulence.
The above frequencies are given by~\cite{Caprini:2024hue}
\begin{equation}
\begin{aligned}
    f_p &\simeq 0.11 \, H_{*,0} \frac{\beta}{H_*} \, , \\
    f_1^{\text{sw}} &\simeq 0.2  \, H_{*,0}  \, (H_* R_*)^{-1} \, , &
    f_2^{\text{sw}} &\simeq 0.5  \, H_{*,0} \, \Delta_w^{-1} \, (H_* R_*)^{-1} \, , \\
    f_1^{\text{tur}} &\simeq  \frac{\sqrt{3\varepsilon K}}{2 \mathcal{N}} H_{*,0} \, (H_* R_*)^{-1}  \, , &
    f_2^{\text{tur}} &\simeq 2.2 \, H_{*,0} \, (H_* R_*)^{-1}   \, , 
\end{aligned}    
\end{equation}
where $ H_{*,0}$ is the Hubble rate at the time of GW production $H_*$ redshifted to today,
\begin{align}
    H_{*,0}=\frac{a_*}{a_0}H_*\simeq1.65\times 10^{-5} \, \textrm{Hz} \, \left(\frac{g_*}{100}\right)^{1/6}\left(\frac{T_*}{100 \,\textrm{GeV}}\right) \, ,
\end{align}
$\Delta_w^{-1}$ is the fluid shell thickness
\begin{align}
    \Delta_w^{-1} = \frac{|v_w-c_s|}{\max(v_w,c_s)} \, ,
\end{align}
with $c_s$ the speed of sound in the plasma, and $\mathcal{N}\simeq 2$ is the number of eddy turnovers.

The power spectra depend on the fractional energy densities
\begin{align}
  \tilde{K} &= \kappa_{\text{col}} \, \alpha/(1+\alpha) \, , 
  &
  K &\approx 0.6 \kappa_{\text{sw}} \, \alpha/(1+\alpha) \, ,
\end{align}
where $\kappa_\text{col,sw}$ are the efficiency factors for the different sources.
In particular, $\kappa_{\text{col}}$ is the fraction of vacuum energy that goes into
accelerating the bubble wall, and $\kappa_\text{sw}$ is the efficiency at which the 
potential energy is converted to fluid kinetic energy~\cite{Athron:2019nbd}.
The factor $0.6$ takes into account the efficiency of producing kinetic energy in the 
bulk fluid motion compared to the single bubble case.
For further details, we refer the reader to 
Refs.~\cite{Caprini:2024hue, Espinosa:2010hh, Jinno:2022mie,Giese:2020rtr,Giese:2020znk}\footnote{
Refs.~\cite{Giese:2020rtr,Giese:2020znk} extend the computation of the efficiency factor beyond the bag model, covering detonations and deflagration/hybrid scenarios, respectively. This requires knowledge of the speed of sound in both phases, a computation we defer to a later version of \packageName.
}.

For the determination of $\kappa_\text{sw}$, we follow Ref.~\cite{Espinosa:2010hh}, 
where 
\begin{align} \label{eq:kappa_sw_Espinosa}
  \kappa_\text{sw} = 
  \begin{cases}
    \frac{c_s^{11/5}\kappa_A\kappa_B}{(c_s^{11/5}-v_w^{11/5})\kappa_B+v_wc_s^{6/5}\kappa_A} \, , &v_w\le c_s \, ,
    \\
    \kappa_B+(v_w-c_s)\delta\kappa+\frac{(v_w-c_s)^3}{(v_J-c_s)^3}l_\kappa \, , &c_s< v_w < v_J \, ,
    \\
    \frac{(v_J-1)^3v_J^{5/2}v_w^{-5/2}\kappa_C\kappa_D}{[(v_J-1)^3-(v_w-1)^3]v_J^{5/2}\kappa_C+(v_w-1)^3\kappa_D} \, , &v_w\ge v_J \, ,
  \end{cases}
\end{align}
with the following shorthands
\begin{align}
    \kappa_A &\simeq v_w^{6/5} \, \frac{6.9 \alpha}{1.36-0.037\sqrt{\alpha}+\alpha}   \, , 
    &
    \kappa_B &\simeq \frac{\alpha^{2/5}}{0.017+(0.997+\alpha)^{2/5}} \, , 
    \\
    \kappa_C &\simeq \frac{\sqrt{\alpha}}{0.135+\sqrt{0.98+\alpha}}  \, , 
    &
    \kappa_D &\simeq \frac{\alpha}{0.73+0.083\sqrt{\alpha}+\alpha} \, , 
    \\
    \delta\kappa &\simeq -0.9\log\frac{\sqrt{\alpha}}{1+\sqrt{\alpha}}   \, , 
    &
    l_\kappa &\simeq  \kappa_C-\kappa_B-(v_J-c_s)\delta\kappa  \, , 
    \\
    v_J &=  \frac{1}{1+\alpha}\left(c_s+\sqrt{\alpha^2+\frac{2\alpha}{3}}\right)  \, .
\end{align}

The efficiency factor for bubble wall collisions, $\kappa_\text{col}$ can be calculated automatically within the package by providing a list of the particles that acquire a mass by crossing the bubble wall, together with their properties (acquired mass, bosonic or fermionic nature).
The computation of the efficiency factors then follows Refs.~\cite{Ellis:2019oqb,Ellis:2020awk}, where an ultra-relativistic bubble wall velocity is assumed.
First we calculate $\alpha_\infty$ and $\alpha_\text{eq}$ defined as
\begin{align}
  \alpha_\infty &= \frac{T^2}{24\rho_\gamma} \left(\sum_b N_b \Delta m_b^2 + \frac{1}{2} \sum_f N_f \Delta m_f^2\right) \, ,  
  &
  \alpha_\text{eq} &= \frac{T^3}{\rho_\gamma} \sum_b N_b g_b^2 \Delta m_b \, ,
\end{align}
where $N_b$ and $N_f$ are the degrees of freedom of bosonic and fermionic species, and
$\Delta m$ is the difference in masses between the broken and unbroken phases.
In the first line, the sums run over all particles that gain mass during the transition, while in the second line only gauge bosons are included, with $g_b$ their gauge couplings.
From these expressions, we compute
\begin{align} \label{eq:gammaeq_gammastar}
  \gamma_\text{eq}&=\frac{\alpha-\alpha_\infty}{\alpha_\text{eq}} \, , 
  &
  \gamma_*&=\frac{2}{3} \frac{R_*}{R_0}\, ,
\end{align} 
where
\begin{equation} \label{eq:HR_from_beta}
    H_*R_*=v_w\, (8\pi)^{1/3}\left(\frac{\beta}{H_*}\right)^{-1} \, ,
\end{equation}
\begin{equation} \label{eq:Rstar_R0}
    R_0=\left( \frac{3\,S_3}{2 \pi \Delta V}\right)^{1/3} \, .
\end{equation}
Eq.~\eqref{eq:HR_from_beta} is a good approximation for fast transitions \cite{PhysRevD.45.3415}\footnote{
In particular, Eq.~\eqref{eq:HR_from_beta} is derived at the e-folding time/temperature $t_e: I_\text{FV}(t_e)=1$. The derivation assumes a linear expansion in time of the action, as is typical for fast transitions, and ignores the Universe expansion \cite{PhysRevD.45.3415}.
}.
Here, $R_0$ is the bubble radius at nucleation, with $\Delta V$ the potential difference between the minima, and $S_3$ the action.
Since, without relying on numerical simulations, tracking the nucleation and expansion history of individual bubbles is not feasible, we evaluate $R_0$ at the nucleation temperature, effectively assuming that the bubbles relevant for this step nucleate around that temperature. Relaxing this assumption lies beyond the scope of our code and existing implementations.
This strategy is often employed in numerical simulations of cosmological phase transitions, see Refs.~\cite{Hindmarsh:2015qta, Hindmarsh:2017gnf, Cutting:2019zws} for an example.
The efficiency factor for bubble wall collisions reads~\cite{Ellis:2019oqb}
\begin{align}
\kappa_\text{col}=
\begin{cases}
    \frac{\gamma_\text{eq}}{\gamma_*}\left[1-\frac{\alpha_\infty}{\alpha}\left( \frac{\gamma_\text{eq}}{\gamma_*}\right)^2\right] \, ,& \gamma_* > \gamma_\text{eq} \, ,
    \\
    1-\frac{\alpha_\infty}{\alpha} \, , & \gamma_* \le \gamma_\text{eq} \, .
    \end{cases}
\end{align}
The efficiency factor for sound waves is then modified to take into account the energy going into the acceleration of the bubble wall instead of into the plasma
\begin{align} \label{eq:kappasw_alt}
\kappa_\text{sw}& =(1-\kappa_\text{col}) \, \frac{\alpha_\text{eff}}{0.73+0.083 \sqrt{\alpha_\text{eff}} + \alpha_\text{eff}} \, , & \alpha_\text{eff}& =\alpha\,(1-\kappa_\text{col}) \, ,
\end{align}
which reduces to the $v_w\ge v_J$ case of \cref{eq:kappa_sw_Espinosa} for $v_w\approx 1$, when $\kappa_\text{col}=0$.
Alternatively, the user has the option to provide a numerical value for $\kappa_\text{col}$ to be inserted into~\cref{eq:kappasw_alt}.
By default, we set the contribution from bubble collisions to $\kappa_\text{col}=0$, as typically this is negligible when there is no runaway behavior.
\section{Download and installation}
\label{sec:download_install_example}
\noindent
To download the package \packageName, please visit the \href{https://github.com/finshky/PT2GW}{dedicated GitHub} page\footnote{
\href{https://github.com/finshky/PT2GW}{https://github.com/finshky/PT2GW}.
}.
To permanently install the paclet, locate the \texttt{.paclet} file and run
\begin{mmaCell}[leftmargin=3.1em,  labelsep=.5em, leftmargin=3em,labelsep=.5em,defined=PacletInstall]{Input}
PacletInstall["/path/to/PT2GW-x.y.z.paclet"]
\end{mmaCell}
where \texttt{x.y.z} is the corresponding version.

\subsection{Quick start} \label{subsec:first_example}
Within the package, a few models are provided as examples. In this section, we show how to utilize the package for the CFF model. For more details, see Section~\ref{subsec:cff_model}.
Once installed, \packageName\ can be loaded by calling
\begin{mmaCell}[leftmargin=3.1em,  labelsep=.5em, leftmargin=3em,
    labelsep=.5em,defined=PT2GW]{Input}
<<PT2GW\`
\end{mmaCell}
\mmaCellGraphics[yoffset=5ex,ig={width=.8\textwidth}]{Print}{mmaGraphics/loadPacletOut}
The CFF model described in Section~\ref{subsec:cff_model} is readily implemented in the \texttt{Models} sub-package as \texttt{CFFModel}:
\begin{mmaCell}[leftmargin=3.1em,  labelsep=.5em, defined={CFFModel,PT2GW,Models,m}]{Input}
<<PT2GW/Models.m
CFFModel[][\mmaUnd{ϕ}, T]
\end{mmaCell}
\begin{mmaCell}[leftmargin=3.1em,  labelsep=.5em]{Output}
1/2 \mmaDef{γ} (T^2 - T0^2) \mmaDef{ϕ}^2 - 1/3 A T \mmaDef{ϕ}^3 + 1/4 \mmaDef{λ} \mmaDef{ϕ}^4
\end{mmaCell}
A few selected benchmark points can be loaded by passing an index argument\footnote{
Since version 1.1.0, loading a benchmark point automatically defines the number of relativistic d.o.f.
}:
\begin{mmaCell}[leftmargin=3.1em,  labelsep=.5em, defined={CFFModel}]{Input}
V = CFFModel[3];
V[\mmaUnd{ϕ}, T]
\end{mmaCell}
\vspace{-3mm}
\includegraphics[trim=-2cm 0 0 0,width=.38\textwidth]{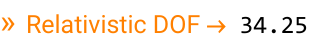}
\vspace{-2mm}
\begin{mmaCell}[leftmargin=3.1em,  labelsep=.5em]{Output}
0.0277778 (-19600. + T^2) \mmaDef{ϕ}^2 
- 0.0146402 T \mmaDef{ϕ}^3 + 0.00385802 \mmaDef{ϕ}^4
\end{mmaCell}

To search for FOPTs in the loaded CFF potential, we simply run \texttt{SearchPotential}: the necessary inputs are the potential $V$ and a numerical value for the bubble wall velocity $v_w$.
Some of the available options include the phase tracing method \texttt{"TracingMethod"}, the options from \texttt{FindBounce} to calculate the bounce profile and action, and the \texttt{"Sources"} of GWs to be included in the power spectrum; see Sections~\ref{sec:method_description} and~\ref{sec:usage_manual} for more details.
\begin{mmaCell}[leftmargin=3.1em,  labelsep=.5em, defined={SearchPotential,V,EchoTiming,NSolve}]{Input}
vw = 0.9;
transitions = SearchPotential[V,vw,
    "TracingMethod"->NSolve];// EchoTiming
\end{mmaCell}

\includegraphics[width=.89\textwidth,
]{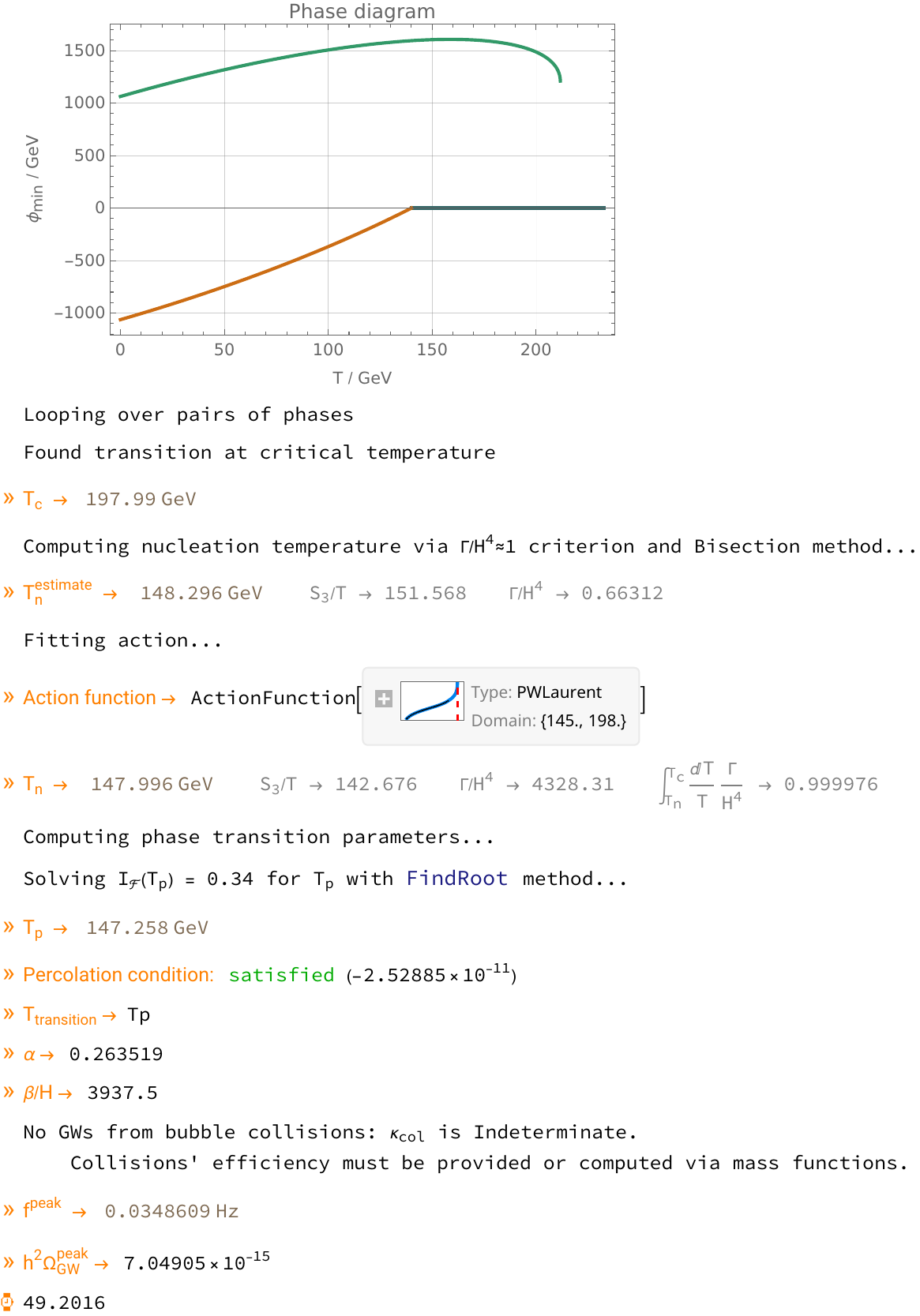}

It is possible for the user to determine the phases independently and provide them to the \texttt{SearchPhases} function, in the form of a \Math\ list of two elements. As an example, to reproduce the previous result one can also run the following:
\begin{mmaCell}[leftmargin=3.1em,  labelsep=.5em, defined={TracePhases,SearchPhases,vw,NSolve,V}]{Input}
Phases = TracePhases[V, "TracingMethod"->NSolve]
tr = SearchPhases[V,Phases[[\{1,3\}]],vw]
\end{mmaCell}

By default, \texttt{SearchPotential} returns a list of \texttt{Transition} objects, containing all the phase transition and SGWB data computed along the way:
\begin{mmaCell}[leftmargin=3.1em,  labelsep=.5em, defined=AutoComplete]{Input}
tr = \mmaDef{transitions}[[1]]
(* optional: enables drop-down key suggestion *)
AutoComplete[tr];
\end{mmaCell}
\mmaCellGraphics[leftmargin=4.1em,labelsep=1em,ig={width=.45\textwidth}]{Output}{mmaGraphics/transition_expanded_cff}
A \texttt{Transition} object is displayed as a \emph{summary box}, presenting the transition strength $\alpha$ and inverse duration $\beta/H$,\footnote{
Data is displayed only if it has been successfully computed.
} which can be expanded to include information about the percolation temperature, and GW spectral peak parameters.
The information contained in a \texttt{Transition} object can be accessed by the corresponding \emph{keys}. For example, to extract the percolation temperature:
\begin{mmaCell}[leftmargin=3.1em,  labelsep=.5em, ,defined=tr]{Input}
tr["Tp"]
\end{mmaCell}
\begin{mmaCell}[leftmargin=3.1em,  labelsep=.5em]{Output}
147.098
\end{mmaCell}
The above \texttt{AutoComplete} function enables us to quickly access the available \emph{keys} from a drop-down menu. Notice how all \packageName\ functions feature autocompletion. Furthermore, \texttt{Transitions} support multiple-keyword search, default values, and function wrapping:
\begin{mmaCell}[leftmargin=3.1em,  labelsep=.5em, defined={tr,Log10},undefined={α,β}]{Input}
tr[\{"fPeak","h2OmegaPeak","not a key"\},"missing",Log10]
\end{mmaCell}
\begin{mmaCell}[leftmargin=3.1em,  labelsep=.5em]{Output}
\{-1.07108, 3.53128, "missing"\}
\end{mmaCell}

A \texttt{Transition} object is essentially a wrapper around an \texttt{Association}.\footnote{
\href{https://reference.wolfram.com/language/guide/Associations.html}{Associations} are the \Math\ equivalent for \emph{dictionaries}.
}
To list all the quantities it stores, it is convenient to display it as a \texttt{Dataset}:
\begin{mmaCell}[leftmargin=3.1em,  labelsep=.5em, defined={tr,Dataset}]{Input}
Dataset[tr]
\end{mmaCell}
\mmaCellGraphics[leftmargin=4.1em,labelsep=1em,ig={width=.45\textwidth}]{Output}{mmaGraphics/dataset_cff}
Furthermore, \packageName\ provides a number of helper functions, listed in Section~\ref{sec:method_description}, to visualize the information contained in a \texttt{Transition}.
It is straightforward to plot diagrams for the phase structure, the Euclidean action, and the SGWB spectrum associated to a \texttt{Transition}.

\begin{mmaCell}[leftmargin=3.1em,  labelsep=.5em, defined={PlotTransition,tr}]{Input}
PlotTransition[tr]
\end{mmaCell}
\mmaCellGraphics[leftmargin=4.1em,labelsep=1em,ig={width=.6\textwidth}]{Output}{mmaGraphics/transition_plot_cff}
Notice the slightly slanted arrow: it indicates a transition ``starting'' at the nucleation temperature, and ``ending'' at the lower percolation temperature. These temperatures are shown explicitly in the plot of the Euclidean action. 
\begin{mmaCell}[leftmargin=3.1em,  labelsep=.5em, morelst={label=cell:Action},defined={PlotAction,tr}]{Input}
PlotAction[tr]
\end{mmaCell}
\mmaCellGraphics[leftmargin=4.1em,labelsep=1em,ig={width=.85\textwidth}]{Output}{mmaGraphics/action_plot_cff}
By default, \texttt{PlotGW} displays the sensitivity curves of a few EW-transition-scale detectors (see Section~\ref{subsec:manual_GW}):
\begin{mmaCell}[leftmargin=3.1em,  labelsep=.5em, defined={PlotGW,tr}]{Input}
PlotGW[tr]
\end{mmaCell}
\mmaCellGraphics[leftmargin=4.1em,labelsep=1em,ig={width=.9\textwidth}]{Output}{mmaGraphics/GW_plot_cff}

\section{Description of the Numerical Methods} \label{sec:method_description}
\noindent
The input required by the package consists of the effective potential and the bubble wall velocity.
Users should construct the effective potential for themselves.
The most common techniques to construct the finite temperature effective potential are daisy resummation~\cite{Parwani:1991gq, Arnold:1992rz} and dimensional reduction~\cite{GINSPARG1980388,PhysRevD.23.2305}.
For the latter, the \texttt{DRalgo} package~\cite{Ekstedt:2022bff} is available, which automatically performs dimensional reduction and constructs the 3-dimensional high-temperature effective theory.
In this case, the \texttt{DRTools} helper (see Section~\ref{subsec:DRTools}) can be used to interface the \texttt{DRalgo} output with the subsequent steps of \packageName.

Within the code, the 3D quantities belonging to the 3D-EFT are rescaled as $V_\text{4D}=T V_\text{3D}$ and $\phi_\text{4D}=\sqrt{T}\phi_\text{3D}$ to compute the thermodynamics of the phase transition.
In this way, all transitions are treated equally and studied in 4 dimensions.

The determination of the bubble wall velocity is beyond the scope of our work, and is set to a predetermined input value. We refer the user to the existing literature on the subject (see for example Refs.~\cite{Moore:1995ua,Moore:1995si,Balaji:2020yrx,Ai:2021kak,Ai:2023see}).
Two package variables can be set by the user after loading the package: the number of relativistic degrees of freedom \texttt{RelativisticDOF}, which defaults to 106.75, i.e. the SM value at the EW scale above the top quark mass, and the units of the problem that are set to GeV, which can be changed with the command \texttt{DefineUnits["GeV"]}.

In Fig.~\ref{fig:package_structure} we outline the general structure of the package, and in the remainder of this Section, we provide an overview on the available routines.
For more details regarding the available options in each routine, we refer the reader to Section~\ref{sec:usage_manual}.


    \tikzstyle{box} = [rectangle, draw, rounded corners, minimum width=3cm, minimum height=1cm, text centered]
    \tikzstyle{subprocess} = [rectangle, draw, rounded corners, minimum width=2.5cm, minimum height=0.8cm, text centered, fill=yellow!40]
    \tikzstyle{searchPhases} = [rectangle, draw, dashed, rounded corners, inner sep=10pt, fit=(Tc) (Tp),
    fill=orange!5,
    label={[anchor=north]above:SearchPhases}]
    \tikzstyle{gw} = [rectangle, draw, rounded corners, inner sep=10pt,
    fill=blue!20]
    \tikzstyle{arrow} = [->, thick]
    
    \tikzstyle{container} = [draw, rectangle, rounded corners, inner sep=10pt]
    
    \def\bottom#1#2{\hbox{\vbox to #1{\vfill\hbox{#2}}}}
    \tikzset{
      mybackground/.style={execute at end picture={
          \begin{scope}[on background layer]
            \node[] at (current bounding box.south){\bottom{1cm} #1};
            \end{scope}
        }},
    }

\begin{figure}
\centering
\scalebox{1}{ 
\begin{tikzpicture}[mybackground={\packageName}]

    \node (pot) [box] {Thermal potential};
    \node (spot) [right=of pot,text opacity=0] {SearchPotential};
    \node (trace) [box, above right=of spot, fill=orange!20] {TracePhases};

    \node (sph) [below=of trace] {SearchPhases};
    \node (Tc) [subprocess, below=0.3cm of sph] {FindCritical};
    \node (TnEst) [subprocess, below=0.3cm of Tc] {$T_n$ estimate};
    \node (action) [subprocess, below=0.3cm of TnEst] {ActionFit};
    \node (Tn) [subprocess, below=0.3cm of action] {FindNucleation};
    \node (Tp) [subprocess, below=0.3cm of Tn] {FindPercolation};
    \node (gw) [gw, right=1.2cm of Tp, label={[shift=(gw.north west)]above right:\texttt{GW}}] {ComputeGW};
    \node (belowTp) [below right=0.1cm of Tp] {};
    \node (fb) [box,fill=red!20,below left=-.3cm and .9cm of TnEst] {FindBounce};
    \begin{scope}[on background layer]
        \node (SearchPotential) [container, fill=green!20, fit=(spot) (trace) (belowTp), label={[anchor=west,xshift=.2cm,yshift=.1cm]above left:SearchPotential}] {};
    \end{scope}
    \begin{scope}[on background layer]
        \node (SearchPhases) [container, fill=orange!20, fit=(sph) (Tp)] {};
    \end{scope}
    
    \node (dr) [box, dashed, above=of pot, label={[shift=(dr.north west)]above right:\texttt{DRTools}}] {DR potential};

    \draw [dashed,->] (dr) -- (pot);
    \draw [arrow] (pot) -- (spot);
    \draw [arrow] (trace) -- (SearchPhases);
    
    \draw [arrow] (Tc) -- (TnEst);
    \draw [arrow] (TnEst) -- (action);
    \draw [arrow] (action) -- (Tn);
    \draw [arrow] (Tn) -- (Tp);
    \draw [arrow,dotted] (fb) -- (TnEst);
    \draw [arrow,dotted] (fb) -- (action);
    
    \draw [arrow] (Tp) -- (gw);

\end{tikzpicture}
}
\caption{Structure of the \packageName\ paclet. The user-defined thermal potential is passed to \texttt{SearchPotential}, which traces its phases and applies the function \texttt{SearchPhases} --- the transition finder --- to every pair of phases. If nucleation is detected, the algorithm fits the action and performs precise calculations of the nucleation and percolation temperatures via Eqs.~\labelcref{eq:Tn_definition,eq:Tp_definition}. \texttt{SearchPhases} computes the fixed-temperature-action using \texttt{FindBounce} \cite{Guada:2020xnz}.
The independent sub-package \texttt{GW} is called to compute the GW background, and it can be adapted to users' preferences. 
The effective potential can be optionally constructed with the dimensional reduction method, by means of \texttt{DRalgo} \cite{Ekstedt:2022bff} and the associated sub-package \texttt{DRTools} we provide. \label{fig:package_structure}
}
\end{figure}


\subsection{Phase tracing and critical temperature}
The default phase tracing method applies the built-in local minimizer \texttt{FindMinimum} to the potential, for a set of temperatures chosen by the user, via a temperature range and a number of sampling points in that interval.
This algorithm assumes there exist, at most, one \emph{broken} and one \emph{symmetric} phase, separated by a threshold in field space defined by the user.
If only one phase is found, the algorithm stops.

Depending on the complexity of the potential, which is a function of the field and the temperature, a semi-analytical method can be applied.
If the potential features a simple field dependence, the semi-analytic tracing methods can be adopted, and three or more phases can be identified.
Each step in a multi-step phase transition is then treated separately.

Once the phases have been determined, the algorithm takes all possible combinations of pairs of phases and checks for an overlap between the temperature intervals where the phases exist.
If an overlap is found, then the critical temperature, at which the minima are degenerate, is calculated using \texttt{FindRoot}, and the analysis of the phase transition proceeds.
If such a temperature is not found, then a different pair of phases is studied.

Our routine allows the user to independently determine the phase structure and provide it to the algorithm searching for phase transitions.
We refer the reader to Section~\ref{sec:usage_manual} for more details on phase tracing.

\subsection{Bounce solution and Euclidean Action}
Once the pair of phases and the relevant temperature range for a transition are identified, the bounce solution and its action are computed over a set of equally-spaced temperatures.
This is done using the \texttt{FindBounce} package~\cite{Guada:2020xnz}, which utilizes the polygonal approach~\cite{Guada:2018jek} to find the bounce solution and the corresponding Euclidean action.
The polygonal method utilizes a segmentation of field space between the two vacua and approximates the potential as a straight line on these segments.
In this way, it is possible to analytically obtain the bounce on each segment and connect consecutive solutions by requiring continuity and derivability. This method can be extended to include perturbations around the linear solution.

For temperature values for which the calculation of the Euclidean action is subject to numerical instabilities, multiple strategies are adopted.
The first is to modify the segmentation of the potential. By default, the two vacua are used as endpoints, but in cases where the exit point is far away from the true vacuum, numerical accuracy is lost because the routine uses a fixed number of segments between the two vacua.
One can first find an approximate value for the exit point, and then run \texttt{FindBounce} utilizing this exit point as an endpoint for the segmentation, instead of the true vacuum, to optimize the segmentation of the potential.
The second is to modify the default \texttt{"ActionTolerance"} setting of \texttt{FindBounce}. We found that a value of $10^{-15}$ generally suffices.
For more details, see Sect.~\ref{subsec:manual_bounceaction}.

The Euclidean action is then fitted. By default, a piecewise function composed of a cubic spline for low temperatures and a Laurent polynomial in powers of $T_c-T$ at high temperatures is adopted:
\begin{align}\label{eq:PWLaurent}
    \left(\frac{S_3}{T}\right)_\text{fit}(T) = \begin{cases}
    \sum_{n=-2}^{1} c_n (T_c-T)^n & T_\mathrm{split}<T<T_c \\
   \mathrm{spline\ interpolation} & T_\mathrm{min}<T\le T_\mathrm{split} \\
    \texttt{Indeterminate} & \mathrm{otherwise}
    \end{cases}\ .
\end{align}
We define the splitting temperature, $T_\mathrm{split}$, as the third-highest temperature value, so that only the three highest-temperature action values are used for the Laurent polynomial fit.
We then impose continuity and match the first and second derivatives at $T_\mathrm{split}$, leaving a single degree of freedom for the Laurent polynomial fit.
The choice of the Laurent polynomial of order -2 is motivated by the expansion of the Euclidean action around $T_c$, see \cref{eq:S3LO_Tc}.
This function captures the positive divergence of the action at the critical temperature.

For simple potentials with a polynomial field-dependence up to fourth order, like the CFF model in Section~\ref{subsec:cff_model}, an analytical form of the Euclidean action can be used; see Section~\ref{subsec:manual_action_analytic} and ~\ref{app:analytics} for more details.

\subsection{Nucleation temperature}
The calculation of the nucleation temperature requires a temperature range as input, together with the overlapping phases and the corresponding Euclidean action for the tunneling between such phases.
Multiple criteria are implemented, such as the integral definition of the nucleation temperature~\cref{eq:Tn_definition} and the approximate criteria~\cref{eq:GammaH-4,eq:ApprCriterionTn}. 
A first estimate can be obtained by a simple bisection algorithm, where the interval is given by the temperature overlap, until the approximate criterion for $T_n$ is satisfied to a predetermined accuracy.
A more precise determination of the nucleation temperature is given by explicitly performing the integral in~\cref{eq:Tn_definition}.
This integral extends all the way to the critical temperature, but the specific details at high temperatures are essentially irrelevant, as the integrand is exponentially suppressed in this region. The goodness of the fit for $S_3/T$ is most important in the region near the nucleation temperature.
If the Euclidean action is known analytically, it is also possible to derive an analytical expression for the nucleation temperature, as explained in \ref{app:analytics}.

The vacuum contribution to the expansion of the Universe might be relevant for some scenarios and is included in the Hubble parameter by default. This is usually important for supercooled transitions, where the nucleation temperature is significantly smaller than the critical temperature.

\subsection{Percolation temperature}
For the calculation of the percolation temperature, the condition $I(T)=0.34$ from~\cref{eq:I_percolation} has to be satisfied. This equation can be solved with the built-in \texttt{FindRoot} or with a bisection algorithm.
For the former, an initial guess for the percolation temperature is needed, which is set to be the nucleation temperature.
In the latter case, the same temperature range as for the calculation of the nucleation temperature is used.
We found it useful to take the logarithm of both sides of~\cref{eq:I_percolation}, in order to mitigate the strong dependence on the temperature, where the integral varies by orders of magnitudes for a small temperature difference.
\subsection{Phase transition parameters}
The phase transition parameters are simply computed by using \cref{eq:alpha} for the strength, and \cref{eq:beta} for the inverse duration, both evaluated at the percolation temperature, by default.
\subsection{Efficiency factors and Gravitational Wave spectrum}
As explained in Section~\ref{subsec:GW_spectrum}, the efficiency factor for bubble wall collisions $\kappa_{\text{col}}$ is set to \texttt{Indeterminate} by default, meaning no bubble wall contribution to the SGWB is considered. In contrast, the efficiency of GW production from sound waves follows the prescription in Ref.~\cite{Espinosa:2010hh}.
In order to calculate the efficiency for bubble collisions, the user has to input the acquired masses of particles that cross the bubble walls. The effective masses must be computed separately.
If the user opts for dimensional reduction, these can be accessed through the \texttt{DRalgo} output.
In particular, the output of \texttt{PrintTensorVEV[]} includes the effective scalar and bosonic mass matrices. For more details, please see Ref.~\cite{Ekstedt:2022bff}.

The templates for the three different sources are adopted from the LISA Cosmology Working Group~\cite{Caprini:2024hue}.
The algorithm outputs the peak amplitude and frequency of the combined spectrum from the three sources (two, if the contribution from bubble wall collision is not computed).
The plotting function for the GW power spectrum allows to include the projected peak-integrated sensitivity curves (PISCs) for LISA, DECIGO and BBO, as well as the power-law-integrated sensitivity curves (PLISCs) detailed in Section~\ref{subsec:manual_GW}, both provided in Ref.~\cite{Schmitz:2020syl}.
By construction, PLISCs do not provide information on the signal-to-noise ratio when the GW spectrum deviates from a power law.
As a result, in realistic situations, they should be considered as a qualitative graphical tool to identify an experiment’s sensitivity.
On the other hand, PISCs work for an arbitrary signal shape, and encode information on the signal-to-noise ratio, which corresponds to the vertical separation between the datapoint and the curve.
They are, however, different for each source of GWs (bubble collisions, sound waves and turbulence).

The calculation of the efficiency factors and the GW power spectrum are contained in a separate file, named \texttt{GW}, that directly communicates with the main phase transition analyzer in \packageName.
It is possible for the user to implement custom templates for both the efficiencies and the power spectra by manually modifying the \texttt{GW} file.

\subsection{Helper function: \texttt{DRTools}}
The \texttt{DRTools} helper module is meant to assist in the transition from the \texttt{DRalgo} output to the needed input for \packageName.
The relevant quantities coming from the dimensional reduction step, such as the $\beta$-functions, the 3D soft and ultrasoft (US) expressions, and the effective potential, are part of the \texttt{DRExpression} variable, in the form of associations.
The user can then use the \texttt{DRStep} function to extract the couplings, masses and other parameters in the 3D effective theory, and construct the analytic effective potential at the desired order with the \texttt{ComputeDRPotentialN} function.
For more details on \texttt{DRTools} we refer to Section~\ref{subsec:DRTools}.
An example demonstrating how to use the helper module is provided in the notebook where the study of the Dark Abelian Higgs model~\cite{Breitbach:2018ddu} is presented (see Section~\ref{subsec:ah_model}).

\section{Examples}\label{sec:examples}
\noindent
In this Section, we present the results of our analysis within the framework of \packageName\ for two different models.
The CFF model features a quartic polynomial potential, so we are able to perform a study based on the fully analytical action obtained in the thin-wall approximation (see \ref{app:analytics}).
We also present numerical results from the direct application of our package.
For the Dark Abelian Higgs model, which features a complex scalar field charged under a $U(1)'$ gauge group, we first perform an analysis based on the daisy resummation of IR divergences, and then using dimensional reduction, showcasing the \texttt{DRTools} helper file to connect the output of \texttt{DRalgo} with \packageName.

\subsection{Coupled fluid-scalar field model}\label{subsec:cff_model}
The CFF model consists of a single real scalar with a quartic polynomial potential, where the temperature-dependent coefficients parametrize the interactions with the ambient cosmic fluid.
This model is widely used for numerical simulations of cosmological phase transitions, and is the perfect example for testing \packageName.
In the following, we compare numerical results from our package (and analytical results) with those of~\cite{Hindmarsh:2015qta} literature, for a series of relevant benchmark points.
The simplicity of the model allows for the use of the analytical results obtained in the thin-wall approximation, which can be applied to obtain the Euclidean action and nucleation temperature in closed form (see \ref{app:analytics} for details).

\begin{figure}[h]
  \centering
  \begin{minipage}[b]{0.49\textwidth}
    \includegraphics[width=\textwidth]{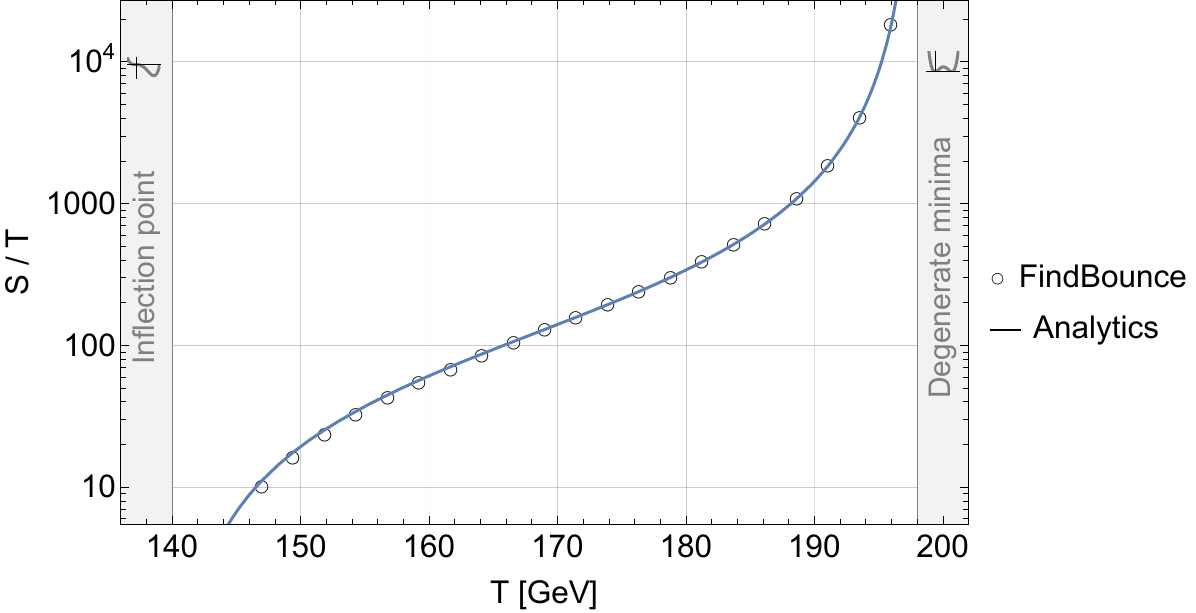}
  \end{minipage}
  \begin{minipage}[b]{0.49\textwidth}
    \includegraphics[width=\textwidth]{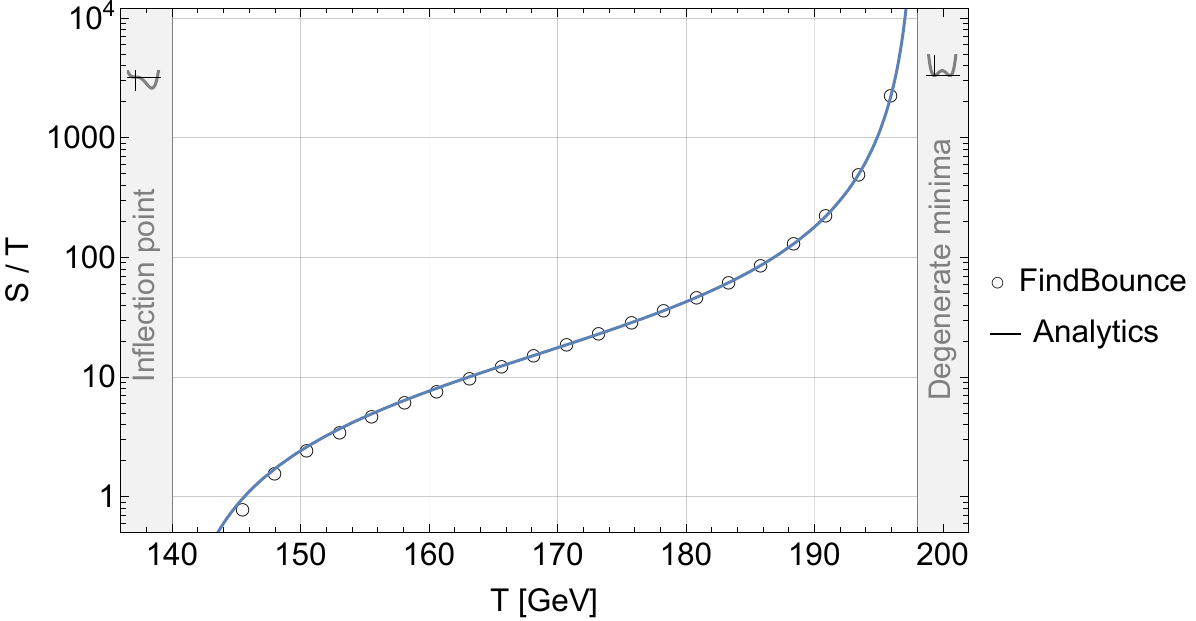}
  \end{minipage}
   \caption{Action calculated analytically (solid line), compared to the numerical result computed via \texttt{FindBounce}, for benchmark points 1 (left) and 2 (right).}
   \label{fig:CFF_Action_analytic_numeric}
\end{figure}

The potential for the CFF model is given by
\begin{equation} \label{eq:VCFF}
    V(\phi,T)=\frac{1}{2}\gamma(T^2-T_0^2)\phi^2-\frac{1}{3}AT\phi^3+\frac{1}{4}\lambda\phi^4 \, .
\end{equation}
The critical temperature is
\begin{equation}
T_c=\frac{\sqrt{9\gamma\lambda}}{\sqrt{9\gamma\lambda-2A^2}} \, T_0 \, ,
\end{equation}
while the inflection point, where the false vacuum and the maximum merge, and only the true vacuum is left, is reached at the temperature $T_0$.
This potential can be directly mapped to the thin-wall potential~\cref{eq:generic_potential} via
\begin{equation}
m^2=\gamma\left(T^2-T_0^2\right) \, , \qquad  \eta=-\frac{1}{3}AT \, , \qquad  \lambda_{C}=2\lambda \, , \qquad \epsal=1-\frac{18\lambda\gamma\left(T^2-T_0^2\right)}{4A^2T^2} \, ,
\end{equation}
which allows us to obtain analytical results for the Euclidean action and the nucleation temperature, as explained in \ref{app:analytics}.
In the following, we will study two benchmark points from Ref.~\cite{Hindmarsh:2015qta}, and an extra benchmark point that is in the remaining parameter space, when the transition happens close to the inflection point.
The benchmark points are summarized in Table~\ref{tab:CFF_benchmarks}, and they are included in the examples within the package (see Section~\ref{subsec:first_example} for details on how to access them).
\begin{table}[ht]
\centering
\begin{tabular}{ |c|c|c|c|c|c|} 
 \hline
 Benchmark & $\gamma$ & $A$ & $\lambda$ &  \, $T_0$ [GeV] &  \, $T_c$ [GeV]\\ 
 \hline
 1 & 1/18 & $\sqrt{10}/72$ & 10/648 & 140 & 197.99\\ 
 2 & 2/9 & $\sqrt{10}/9$ & 160/648 & 140 & 197.99\\
 3 & 1/9 & $\sqrt{10}/144$ & 1.25/648 & 140 & 197.99\\
 \hline
\end{tabular}
\caption{Benchmark points for the study of the CFF model. Benchmark points 1 and 2 are taken from Ref.~\cite{Hindmarsh:2015qta}.
}
\label{tab:CFF_benchmarks}
\end{table}
In Fig.~\ref{fig:CFF_Action_analytic_numeric} we compare the action obtained analytically with the numerical results obtained via \texttt{FindBounce}, for benchmark points 1 and 2.
We see that there is excellent agreement over the whole range of temperatures, which leads to consistent results between analytics and numerics.

In Table~\ref{tab:CFF_comparison_with_literature} we compare the nucleation temperature and the phase transition strength from Ref.~\cite{Hindmarsh:2015qta} with our results from the analytical Euclidean action, and from \packageName\ using the numerical action from \texttt{FindBounce}.
Note that in Ref.~\cite{Hindmarsh:2015qta} the nucleation temperature is set by hand to be $T_n=0.86\, T_c$, as its precise determination was not the focus of that work.
Nonetheless, the strength of the transition is comparable with our result, as the dependence on the temperature is not strong.
\begin{table}[ht]
\centering
\begin{tabular}{|c| ccc| ccc|}
\hline
\multirow{2}{*}{Benchmark} & \multicolumn{3}{c}{$T_n$ [GeV]} & \multicolumn{3}{c|}{\(\alpha_{T_n}\)} \\
\cline{2-4} \cline{5-7}
 & Estimate~\cite{Hindmarsh:2015qta} & Analytic & Numeric & Estimate~\cite{Hindmarsh:2015qta} & Analytic & Numeric \\
\hline
1 & $0.86\,T_c=170.28$ & 170.02 & 170.26 & 0.010 & 0.010 & 0.010 \\
2 & $0.86\,T_c=170.28$ & 188.83 & 188.89 & 0.010 & 0.009 & 0.009\\
 3 & /          &   147.52 & 148.03 & / & 0.331 & 0.331 \\
\hline
\end{tabular}
\caption{Nucleation temperature and phase transition strength from Ref.~\cite{Hindmarsh:2015qta}, from the action calculated analytically, and from the action calculated numerically. Note that the definition of strength in Ref.~\cite{Hindmarsh:2015qta} is equivalent to \cref{eq:alpha} in the CFF model.}
\label{tab:CFF_comparison_with_literature}
\end{table}

On the numerical side, \packageName\ takes about 45 seconds to analyze each benchmark point.
In Table~\ref{tab:CFF_package} we summarize our results.
We used $g_*=34.25$ (in accordance with Ref.~\cite{Hindmarsh:2013xza}), $v_w=0.95$ and $\kappa_{col}=0$, i.e. no gravitational wave production from bubble collisions.
\begin{table}[ht]
\centering
\begin{tabular}{ |c|c|c|c|c|c|c|c| } 
 \hline
 Benchmark & $T_c$ & $T_n$ & $T_p$ & $\alpha$ & $\beta/H$ & $f_{GW}^\text{peak}$  [Hz] & $h^2 \Omega_{GW}^\text{peak}$ \\ 
 \hline
 1 & 197.99 & 170.26 & 168.60 & $4.80\times10^{-3}$ & $1.70\times10^{3}$ & $1.42\times10^{-2}$ & $4.74\times10^{-19}$ \\ 
 2 & 197.99 & 188.89 & 188.03 & $2.93\times10^{-3}$ & $4.02\times10^{3}$ &  $3.72\times10^{-2}$ & $1.96\times10^{-20}$ \\
 3 & 197.99 & 148.03 & 147.18 & $2.64\times10^{-1}$ & $3.19\times10^{3}$ & $2.45\times10^{-2}$ & $9.66\times10^{-15}$  \\
 \hline
\end{tabular}
\caption{Numerical results for the relevant temperatures, phase transition parameters, peak frequency and amplitude. Note that here we use the definition of the PT strength given by~\cref{eq:alpha}. Temperatures are in GeV.}
\label{tab:CFF_package}
\end{table}

\begin{figure}[h]
    \centering
    \includegraphics[width=\linewidth]{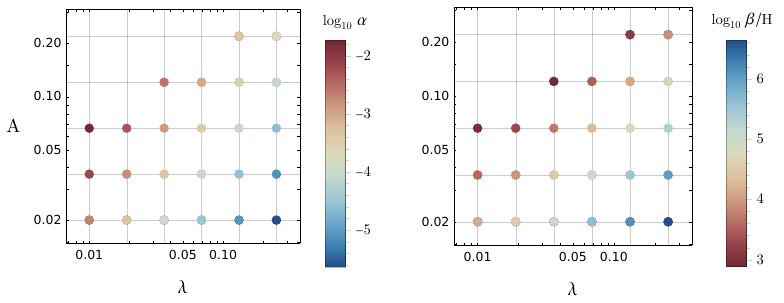}
    \caption{A grid of phase transitions identified in the CFF model. 
    Color indicates values of the strength $\alpha$ on the left panel, and of the inverse duration $\beta/H$ on the right panel.
    }\label{fig:CFF_scan}
\end{figure}

In Fig.~\ref{fig:CFF_scan} we illustrate a scan of the $(\lambda,A)$ parameter space of the model, where we fixed $T_0=140~\text{GeV}$ and $\gamma=1/9$. The data and plots may be reproduced with the \texttt{CoupledFluidField} example notebook.
For fixed cubic coupling $A$, a smaller quartic coupling $\lambda$ leads to a stronger and longer transition: this is due to the fact that a small $\lambda$ corresponds to a deeper broken minimum.
For fixed $\lambda$, a growing $A$ corresponds to a larger potential barrier, and the transition is strong and slow.
For even larger potential barriers, the field is trapped in the FV, and there is no transition.

\subsection{Dark Abelian Higgs}\label{subsec:ah_model}
As a second testing ground for our package, we consider a model consisting of a dark complex scalar $S$ that is a singlet under the SM, but is charged under a dark $U(1)'$.
The terms in the Lagrangian involving the Higgs field $H$ and the dark particles $S$ and $A_\mu'$ are 
\begin{equation}
    \mathcal{L} \supset |D_\mu S|^2 
    + |D_\mu H|^2
    -\frac{1}{4}F_{\mu\nu}'F'^{\mu\nu}
    -\frac{\delta}{2}F_{\mu\nu}'F^{\mu\nu}
    -V(S,H) \, ,
\end{equation}
where $F_{\mu\nu}'$ and $F_{\mu\nu}$ are the field strength tensors of $U(1)'$ and $U(1)_Y$, respectively, and the covariant derivative on $S$ is $D_\mu S=(\partial_\mu+ig_DA_\mu')S$, with $g_D$ being the dark gauge coupling.
The tree level potential is given by 
\begin{equation}
    V_\text{tree}(S,H)=-\mu_S^2S^\dagger S -\mu^2H^\dagger H+\frac{\lambda_S}{2}(S^\dagger S)^2+\frac{\lambda}{2}(H^\dagger H)^2 +\lambda_\text{SH}(S^\dagger S)(H^\dagger H).
\end{equation}
In particular, we consider the model presented in Ref.~\cite{Breitbach:2018ddu}, where the dark sector is completely secluded, i.e.\ $\delta=0$ and $\lambda_\text{SH}=0$.
The theory then reduces to a single-field model involving only the dark scalar $S$, with tree level potential
\begin{align} \label{eq:Vtree_darkAH}
    V_\text{tree}(S)=-\mu_S^2S^\dagger S + \frac{\lambda_S}{2}(S^\dagger S)^2 \, .
\end{align}
The tree-level minimization condition for the scalar potential is
\begin{equation} \label{eq:treelevel_min}
    \mu_\text{S}^2 = \frac{\lambda_\text{S}}{2}\left(v_\text{S}^0\right)^2 \, .
\end{equation}
We first utilize the daisy resummation approach to compute the effective potential, feed it into our code, and compare results with some benchmark points from Ref.~\cite{Breitbach:2018ddu}.
We then perform dimensional reduction using \texttt{DRalgo}, to compare the results with the daisy resummation approach and to give an example of the use of the \texttt{DRTools} helper.
Notice that for this model, the temperature of the dark sector does not have to be the same as the SM temperature.
In this Section, we denote the hidden sector temperature as $T$, and the ratio of temperatures is given by
\begin{align}
    \xi\equiv\frac{T}{T_\gamma} \, ,
\end{align}
where $T_\gamma$ is the photon temperature.
We will set $\xi=0.48$, in accord with some of the benchmark points studied in Ref.~\cite{Breitbach:2018ddu}.
This affects the radiation energy density
\begin{align}
    \rho_\gamma=\frac{\pi^2}{30}\left(\frac{g_{*\text{,SM}}}{\xi^4}+g_h\right)T^4 \, ,
\end{align}
and, in turn, the Hubble parameter, which during radiation domination reads
\begin{align}
    H^2=\frac{\rho_\gamma}{3M_\text{PL}^2} \, .
\end{align}
Here $g_{*\text{,SM}}$ and $g_h$ are the numbers of relativistic degrees of freedom in the SM and in the hidden sector, respectively.
For the model at hand, $g_h=5$.
Moreover, given we will study transitions at the keV scale, we consider only photons and neutrinos in the SM bath, for which $g_{*\text{,SM}}\approx3.38$~\cite{Baumann:2022mni}, giving a total number of $g_*\approx68.67$  relativistic degrees of freedom.

\subsubsection{Daisy resummation approach}\label{subsubsec:dp_daisy_resummation}
The zero-temperature renormalized Coleman-Weinberg potential is
\begin{equation}
    V_\text{CW}(S)=\sum_i \frac{\eta_i n_i}{64\pi^2}m_i^4(S)\left[\log\left(\frac{m_i^2(S)}{\Lambda^2}\right)-C_i\right]+V_\text{ct}(S) \, ,
\end{equation}
where $i$ is an index for the particles with a $S$-dependent mass, $\eta_i=1$ for bosons and $\eta_i=-1$ for fermions, $n_i$ is the number of degrees of freedom for the particle, and $C_i=3/2$ for scalars and fermions and $C_i=5/6$ for gauge bosons.
In accordance with Ref.~\cite{Breitbach:2018ddu}, the renormalization scale $\Lambda$ is chosen to be the tree-level VEV $v_S^0$.
For this model, the particles entering the sum are
\begin{equation}
\begin{split}
    \text{radial scalar mode: } & m_{\phi_S}^2(S)=-\mu_S^2+\frac{3}{2}\lambda_S S^2 \, , \\
    \text{Goldstone mode: } & m_{\sigma_S}^2(S)=-\mu_S^2+\frac{1}{2}\lambda_S S^2 \, , \\
    \text{gauge boson: } & m_{A'}^2(S)= g_D^2 S^2 \, .
\end{split}
\end{equation}
The counterterm potential is
\begin{equation}
    V_\text{ct}(S)=-\frac{\delta\mu_S^2}{2}S^2+\frac{\delta\lambda_S}{8}S^4 \, ,
\end{equation}
where the counter-terms are determined via the following renormalization conditions
\begin{align}
    \left.\frac{\partial V_\text{CW}(S)}{\partial S}\right|_{S=v_S^0} &\overset{!}{=} 0    \, , \\
    \left.\frac{\partial^2 V_\text{CW}(S)}{\partial S^2}\right|_{S=v_S^0} &\overset{!}{=} 0 \, ,
\end{align}
that fix the zero-temperature VEV and mass to their values at tree-level.

The finite-temperature part of the one-loop potential is given by
\begin{equation}
    V_T(S,T)=\sum_i\frac{\eta_i n_i T^4}{2\pi^2} \int_0^\infty \text{d}x \, x^2 \log\left[1-\eta_i \exp\left(-\sqrt{x^2+m_i^2(S)/T^2}\right)\right] \, .
\end{equation}
A high-temperature and a low-temperature expansions are available, but we will not use them in the following.
The theory is ill-behaved in the IR, due to an infinite series of diverging diagrams, the daisy diagrams, where the IR modes are screened by the UV modes.
To tackle this divergence, a resummation is employed~\cite{Arnold:1992rz}, which gives rise to the following contribution to the effective potential
\begin{align}
    V_\text{daisy}(S,T)=-\frac{T}{12\pi}\sum_\text{bosons}n_i \left[\left(m^2(S)+\Pi(T)\right)_i^{3/2}-\left(m^2(S)\right)_i^{3/2}\right] \, .
\end{align}
Here $\Pi(T)$ is the temperature-dependent Debye mass, which vanishes for the transverse gauge boson modes, while for the longitudinal gauge boson mode and for the scalar modes is given by
\begin{align}
    \Pi_{A'}(T) &= \frac{1}{3}g_D^2 T^2 \, , \\
    \Pi_S(T) &= \left(\frac{\lambda_S}{6}+\frac{g_D^2}{4}\right)T^2 \, ,
\end{align}
respectively.
The full effective potential $V(S,T)=V_\text{tree}(S)+V_\text{CW}(S)+ V_T(S,T)+V_\text{daisy}(S,T)$ is then fed into our package.
Within the \texttt{Examples} folder in \packageName, we provide the \texttt{DarkAbelianHiggs\_CW} notebook that reproduces this analysis.

\begin{figure}[t]
    \centering
    \includegraphics[height=0.4\linewidth]{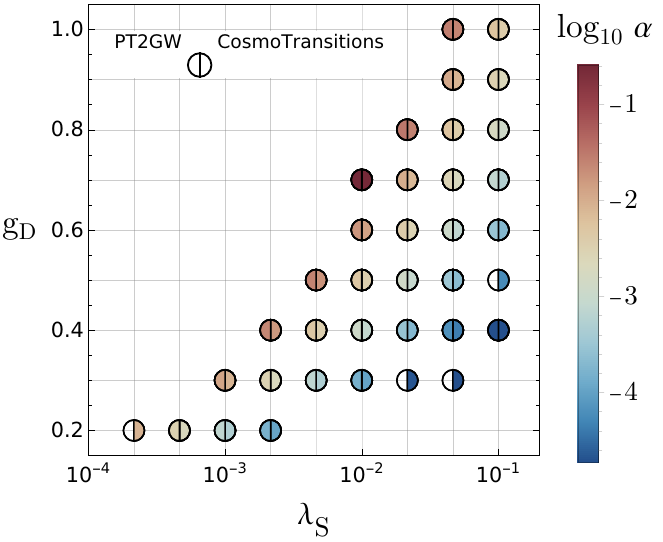}
    \hfill
    \includegraphics[height=0.4\linewidth]{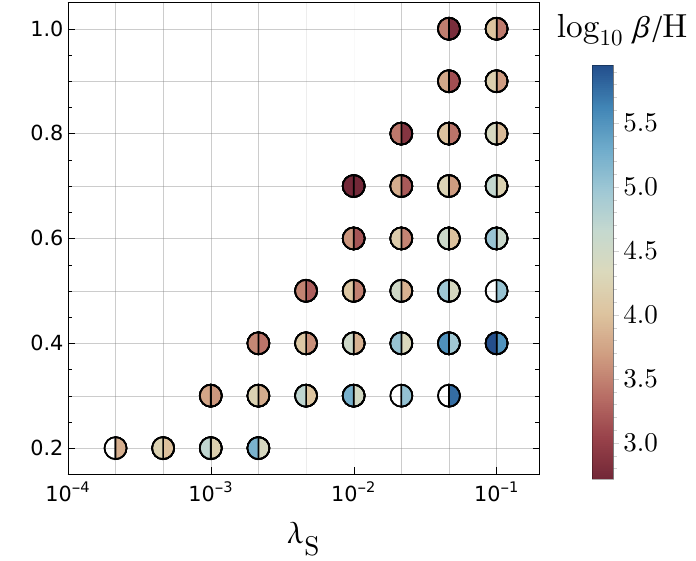}
    \caption{Comparison between \packageName\ (left half-circles) and \texttt{CosmoTransitions} (right half-circles) results on the transitions identified in the Dark Abelian Higgs model, with daisy resummation.
    At the fixed tree-level VEV value $v_S^0=40~\mathrm{keV}$, we scan on a grid of values for the quartic coupling $\lambda_S$ and gauge coupling $g_D$.
    Color indicates values of the strength $\alpha$ on the left panel, and of the inverse duration $\beta/H$ on the right panel.
    The analysis is designed to partially replicate Fig.\ 8 of Ref.~\cite{Breitbach:2018ddu}.}\label{fig:darkPhoton_compare}
\end{figure}

In Fig.~\ref{fig:darkPhoton_compare} we report our results.
The left-half circles are obtained with our \packageName\ package, while the right-half circles adopt \texttt{CosmoTransitions}.\footnote{
    We thank the authors of Ref.~\cite{Breitbach:2018ddu} for sharing their \texttt{CosmoTransitions} implementation of the Dark Abelian Higgs model, which we adopted to construct the effective potential and run the search for FOPTs.
    We then implemented our own modules to obtain the phase transition parameters presented in Fig.~\ref{fig:darkPhoton_compare}.
}
In analogy with Fig.~8 of Ref.~\cite{Breitbach:2018ddu}, we identify several transitions at a fixed tree-level-VEV of $v_S^0=40~\mathrm{keV}$.
The bubble wall velocity is set to $v_w=0.99$.
While the authors of Ref.~\cite{Breitbach:2018ddu} compute $\alpha$ and $\beta/H$ at the nucleation temperature, the transition parameters of Fig.~\ref{fig:darkPhoton_compare} are derived at the percolation temperature, in both the \packageName\ and \texttt{CosmoTransitions} implementations.
Adopting the built-in \Math\ parallelization, we obtained the results of Fig.~\ref{fig:darkPhoton_compare} in $\sim6~\text{m}~40~\text{s}$, on a \texttt{x86-64} processor with 14 kernels.

The size of the barrier is controlled by $g_\text{D}$ via the one-loop cubic term in the effective potential.
For fixed $g_D$, the potential energy difference increases as $\lambda_S$ decreases, leading to stronger and slower transitions.
In the lower-right region of the parameter space, for large $\lambda_\text{S}$ and small $g_\text{D}$, the transition is a crossover.
When a FOPT is present, the transition slows down and the strength increases as $g_\text{D}$ grows. For even larger values of $g_\text{D}$ the scalar VEV remains trapped.
Along diagonals (from lower-left to upper-right) in the parameter space, we observe that the phase transition parameters remain approximately constant. 
This is explained by the competing behaviors of $g_D$ and $\lambda_S$: as we increase both, the potential barrier rises, but at the same time the broken minimum deepens.

The results are generally consistent between the \packageName\ and the \texttt{CosmoTransitions} implementations, for both the phase transition strength and duration.
Although we use similar methods to compute $\alpha$ and $\beta/H$, the fitting of the Euclidean action as a function of temperature is less controlled in our Python implementation based on \texttt{CosmoTransitions}, leading to larger deviations in $\beta/H$, which is particularly sensitive to the temperature dependence of the action.

The \packageName\ package is unable to identify a few weak, fast transitions, in the lower-right end of the parameter space.
In the lower-left corner, corresponding small values of both $g_\text{D}$ and $\lambda_\text{S}$, the potential becomes very flat as the temperature approaches the inflection point.
In the lower-left corner of Fig.~4, corresponding to very small values of both $g_\text{D}$ and $\lambda_\text{S}$, the potential becomes very flat as the temperature approaches the inflection point.
In the automated scan used to produce Fig.~4, we adopt a fixed set of options and hyperparameters; with this choice, our bounce solver is unable to compute the Euclidean action for small $g_\text{D}$ and $\lambda_\text{S}$, even with the \texttt{RefineInflection} function (see Section~\ref{subsec:action_filter_refine}), and the nucleation temperature cannot be determined. By choosing an appropriate shift parameter $s$ for the estimated exit point of the bounce (see \cref{sec:exit_point}), the polygonal bounce method successfully captures this transition. With this adjustment, the action can be computed over a temperature range including nucleation and percolation, and the resulting transition parameters agree with those obtained using \texttt{CosmoTransitions}. An automated procedure to determine the optimal shift dynamically will be included in a future \packageName\ release.

\subsubsection{Dimensional reduction approach} \label{sec:dimensionalReduction}
We perform an additional analysis by computing the effective potential in the dimensional reduction approach, using the \texttt{DRalgo} package. More specifically, we match~\cref{eq:Vtree_darkAH} to a 3D effective potential at the \emph{soft} scale, where the matching is performed at NLO. In order to make a reasonable comparison to the daisy resummation approach in Section~\ref{subsubsec:dp_daisy_resummation}, we derive the potential at LO.

To derive the effective potential in a closed form, i.e. a function of the field and temperature values of the form $V(\phi,T)$, we use the \texttt{DRTools} sub-package, available to the user.
Given the expressions obtained with \texttt{DRalgo}, \texttt{DRTools} automatizes the following steps:
\begin{outline}
    \1 solve numerically the Renormalization Group (RG) equations;
    \1 derive the effective, temperature-dependent parameters (including scalar and vector masses, Debye masses, cubic and quartic couplings);
    \1 construct the effective potential as a function of the effective parameters: $V_{\text{3D}}(\phi_{\text{3D}})$;
    \1 rescale to a 4D theory, according to
   \begin{align} \label{eq:rescale_to_4D}
    &V_{\text{4D}}    = T V_{\text{3D}} \, , &\phi_{\text{4D}} = \sqrt{T}\phi_{\text{3D}} \, . 
\end{align}
    This is an optional step, performed by default.
\end{outline}
\begin{figure}[t]
    \centering
    \includegraphics[width=0.48\linewidth]{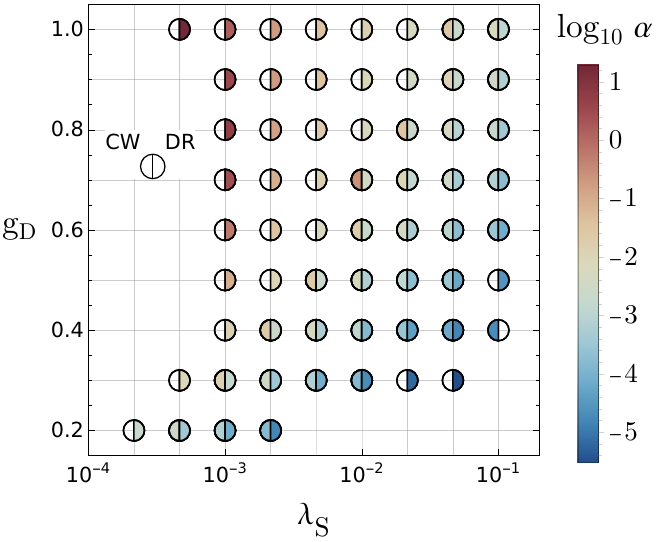}
    \hfill
    \includegraphics[width=0.48\linewidth]{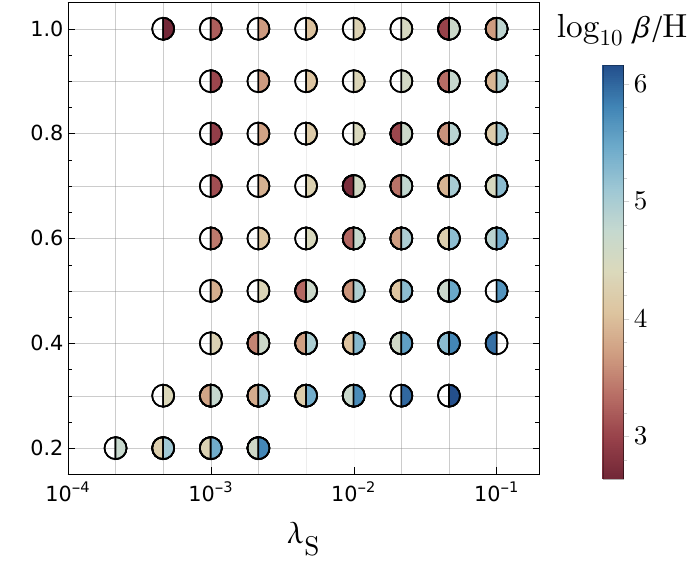}
    \caption{Comparison between the phase transitions identified for the Dark Abelian Higgs model, with the effective potential computed in the Coleman-Weinberg (\texttt{CW}, left half-circles) and dimensional reduction approaches (\texttt{DR}, right half-circles) respectively.
    The benchmark points tested are the same as in Fig.~\ref{fig:darkPhoton_compare}. 
    Color indicates values of the strength $\alpha$ on the left panel, and of the inverse duration $\beta/H$ on the right panel.
    The analysis is designed to partially replicate Fig.\ 8 of Ref.~\cite{Breitbach:2018ddu}.}\label{fig:darkPhotonDR_compare}
\end{figure}
In Fig.~\ref{fig:darkPhotonDR_compare} we compare two analyses made using our package, one with the daisy resummed potential --- i.e. \emph{Coleman-Weinberg} potential --- that corresponds to the results already shown in Section~\ref{subsubsec:dp_daisy_resummation}, and one with the dimensionally reduced potential as explained above. We obtained the results in the dimensionally reduced approach in $\sim5~\text{m}~47~\text{s}$ (see the \texttt{DarkAbelianHiggs\_DR} notebook in the \texttt{Examples} folder of \packageName).
The results are relatively consistent between the two methodologies, but it is worth noticing that the dimensional reduction approach generally leads to a slightly stronger and slower transition, for a given pair of $(\lambda_\text{S}, \, g_\text{D})$.
We notice that the parameter space in which a FOPT is present is much more extended than before, with large values of $g_\text{D}$ corresponding to very strong and slow transitions.
The reason lies in the fact that the effective potential is qualitatively different.
In the CW implementation, the broken minima have higher potential energy, and no critical temperature exists. In the DR implementation, however, the phase structure is such that a FOPT from the symmetric minimum at the origin to the broken minimum is possible.

\section{User's Manual} \label{sec:usage_manual}
\noindent
This section details the available options for each function. Following \Math's \emph{rule} construct, we report the default value for each option with a right arrow: \texttt{<option name>} $\rightarrow$ \texttt{<default value>}.

\subsection{Phase tracing}
The \texttt{TracePhases[V]} function requires the potential as input, and allows for the following options:
\begin{outline}
    \1 \texttt{"TracingMethod"} $\rightarrow$ \texttt{"Numeric"}: sets the method used to trace the phases. We list the various options, arranged from a semi-analytic to a purely numeric approach.
    Semi-analytical methods provide more analytical control and are faster, as they avoid repeated numerical minimizations with \texttt{FindRoot}/\texttt{FindMinimum}.
    Nevertheless, we find that the \texttt{"Numeric"} method remains relatively fast and reliable in practice.
    Users can also perform phase tracing separately and feed the phases to \texttt{SearchPhases} if desired.
        \2 \texttt{NSolve}: the built-in \texttt{NSolve} function can be used to solve the system
        \begin{align}\label{eq:minimizePotential}
        \begin{cases}
            \partial_\phi V(\phi,T) = 0 \\
            \partial^2_\phi V(\phi,T) > 0
            \end{cases}
        \end{align}
        analytically, when the potential parameters are numerical. This method applies when the dependence of the potential on both the field and temperature are reasonably simple;
        \2 \texttt{"SimpleFieldDependence"}: for potentials with a simple field dependence but more involved temperature dependence,\footnote{
        For example, the effective potentials obtained via dimensional reduction (see Section~\ref{sec:dimensionalReduction}) typically feature non-analytic, interpolating functions of the temperature, as a result of the RG evolution of the parameters.
        } we iteratively perform the minimization at fixed temperature. With this option, we once again use \texttt{NSolve} to solve~\cref{eq:minimizePotential}, but we implement this on a list of fixed temperatures ${T_i}$, and thereafter interpolate between the solutions to construct the phases.
        Therefore, this method requires the specification of a temperature range, which is uniformly sampled by the number of tracing points (see the \texttt{"TRange"} and \texttt{"NTracingPoints"} options below), together with a threshold in field space (see \texttt{"SymmetricPhaseThreshold"} option below) that separates the regions in which the algorithm looks for a numerical minimization of the potential using the built-in \texttt{NSolve} method.
        \2 \texttt{"Numeric"} (default): if the field dependence of the potential is more involved, we cannot solve~\cref{eq:minimizePotential}. We implement the built-in local minimizer \texttt{FindMinimum} on a set of uniformly-sampled temperatures ${T_i}$, selected as described above.
    \1 \texttt{"TRange"}$\rightarrow$\texttt{None}: selects the temperature range for the \texttt{"SimpleFieldDependence"} and \texttt{"Numeric"} methods;
    \1 \texttt{"NTracingPoints"}$\rightarrow$\texttt{100}: specifies the number of temperature values, at which the potential is minimized within the temperature range for the \texttt{"Numeric"} and \texttt{"SimpleFieldDependence"} methods;
    \1 \texttt{"SymmetricPhaseThreshold"}$\rightarrow$\texttt{0.1} (in units of energy): specifies the threshold separating the regions in field space where the algorithm looks for a symmetric phase ($\abs{\phi_S}\le\phi_\text{thr}$) and a broken phase ($\abs{\phi_B}>\phi_\text{thr}$). We advise to modify this value according to the specific model at hand, for example by setting it to be smaller than the VEV of the theory;
    \1 \texttt{"BrokenPhaseScale"}$\rightarrow$\texttt{$10^6$} (in units of energy): for the \texttt{"Numeric"} tracing method, sets the starting point in field space when looking for the broken phase. To avoid falling into the FV basin, it should be chosen to be large. We advise to modify this value according to the specific model at hand, for example by setting it to be much larger than the VEV of the theory;
    \1 \texttt{"ShiftSymmetricPhase"}$\rightarrow$\texttt{False}: in case there are numerical issues with the derivatives of the potential at the origin of field space, it is possible to shift the value of the symmetric phase by a small amount, without affecting the rest of the calculation. Instead of using this option, the user is advised to redefine the potential by taking the limit of the field going to zero when necessary;
    \1 \texttt{"PlotPhaseDiagram"}$\rightarrow$\texttt{True}: returns the plot of the phases as a function of temperature;
    \1 \texttt{ProgressIndicator}$\rightarrow$\texttt{True}: determines whether to display a progress bar. If the phase tracing is numerical, the progress bar shows the current temperature;
    \1 \texttt{Print}$\rightarrow$\texttt{True}: determines whether to temporarily print a status messages.
\end{outline}
\subsubsection{Phases overlap}
The function \texttt{Overlap[$\phi_1,\phi_2$]} returns the temperature overlap of any pair of phases $\phi_1, \phi_2$. The option \texttt{RegionBounds}, which is set to \texttt{True} by default, gives the bounds for the overlapping region.
\subsection{Critical temperature}
The critical temperature is calculated with the \texttt{FindCritical[V,\{$\phi_1,\phi_2$\}]} function, by taking any pair of phases with non-empty overlap.
Starting from the midpoint of the overlap temperatures interval, a vanishing potential difference between the two minima is found using \texttt{FindRoot}.
A message is returned in the following cases: (i) there is no overlap between any pairs of phases, (ii) no critical temperature is found, or (iii) multiple critical temperatures are found.
\subsection{Bounce action} \label{subsec:manual_bounceaction}
\subsubsection{Euclidean action and \texttt{FindBounce}}
The \texttt{Action[T,V,\{$\phi_1,\phi_2$\}]} function, specifically designed for thermal transitions, serves as a wrapper for the \texttt{FindBounce} package. While it inherits all options from \texttt{FindBounce}, the following options are explicitly modified to comply with thermal phase transitions:
\begin{outline}
    \1 \texttt{"FieldPoints"}$\rightarrow$\texttt{51}: from \texttt{FindBounce}, number of segmentation points in field space between the two minima;
    \1 \texttt{"Dimension"}$\rightarrow$\texttt{3}: from \texttt{FindBounce}, number of dimensions. The default value is relevant for thermal transitions;
    \1 \texttt{"Gradient"}$\rightarrow$\texttt{None}: from \texttt{FindBounce}, determines whether the beyond-the-linear approximation is used for the potential. Other possible options are \texttt{Automatic}, \texttt{"FiniteDifferences"} and a list of pre-computed values \{$\di V/\di\phi_i$\};
\end{outline}
For more details regarding \texttt{FindBounce}, we refer to Ref.~\cite{Guada:2020xnz}.
The options listed below are proper to the \texttt{Action} function of \packageName :
\begin{outline}
    \1 \texttt{"Rebounce"}$\rightarrow$\texttt{\{True,"ShiftToExitPoint"->True,"LowerActionTolerance"->True\}}: determines whether the bounce solution and action are recomputed when \texttt{FindBounce} encounters numerical errors, giving a non-numerical result. As described below, there are two possible ways to circumvent this issue: shifting the exit point or lowering the action tolerance. By default, both are applied, in this order. To suppress either of these methods, set the corresponding option to \texttt{False}. To avoid entirely the recomputation of the action in case of failure, set \texttt{"Rebounce"}$\rightarrow$\texttt{\{False\}};
    \1 \texttt{"CheckProfile"}$\rightarrow$\texttt{\{True,"ShiftToExitPoint"->True,"LowerActionTolerance"->True\}}: similarly to the \texttt{"Rebounce"} option, allows to recompute the action when the bounce profile displays a discontinuity, measured by \texttt{BounceError} (see below), above a fixed threshold;
    \1 \texttt{"PrintBounceInfo"}$\rightarrow$\texttt{False}: outputs info on the bounce solution, including possible errors and solutions of such;
    \1 \texttt{"PrintAction"}$\rightarrow$\texttt{False}: if set to \texttt{True}, prints the value of the action at the specific input temperature;
    \1 \texttt{"PrintShiftInfo"}$\rightarrow$\texttt{False}; if set to \texttt{True}, prints information about the shift of the exit point;
    \1 \texttt{"PlotBounce"}$\rightarrow$\texttt{False}: determines whether to plot the bounce profile;
    \1 \texttt{"Action/T"}$\rightarrow$\texttt{True}: option to decide whether the output should be $S(T)/T$, which is the default option, or $S(T)$.
\end{outline}
The \texttt{BounceError[$\phi_b$]} determines whether the bounce profile $\phi_b$ is accurate enough.
It is possible that \texttt{FindBounce} returns a profile that evaluates as correct, while it is not.
An example we have found is when at one of the \texttt{FindBounce Radii} the bounce solution is discontinuous.
To fix this, we check that the average of the discontinuities at all the $N_r$ \texttt{Radii} is smaller than a predetermined threshold, which defaults to $t=10^{-3}$, i.e.
\begin{align}
    \left(\sum_{i=1}^{N_r}\frac{\left.(\phi_{i+1}-\phi_i)\right|_{r=r_i}}{\left.(\phi_{i+1}+\phi_i)\right|_{r=r_i}}\right)^{1/2}\le t \, .
\end{align}
\subsubsection{Exit point shifting} \label{sec:exit_point}
\begin{figure}
    \centering
    \includegraphics[width=0.6\linewidth]{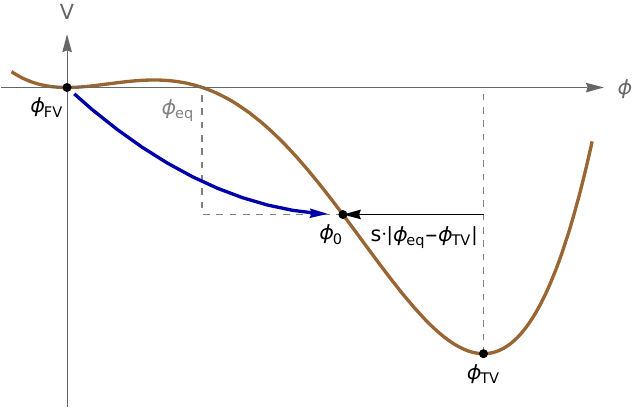}
    \caption{Illustration of the algorithm to shift the field range from the true vacuum $\phi_\text{TV}$ to an estimated \emph{exit point} $\phi_0$. The estimate is given by a shift of an amount proportional to $(\phi_\text{TV}-\phi_\text{eq})$ from the equivalence point $\phi_\text{eq}$. Here, $s$ is the proportionality constant.
    The orange arrow represents the action path in field space, after the shift.
    The blue arrow represents the tunneling path in field space, after the shift.
    }
    \label{fig:shift2ExitPoint}
\end{figure}
While, by default, \texttt{FindBounce} computes the bounce action between two minima of the potential, 
the path of the bounce solution may involve, in practice, only a subsection of the field space.
In other words, instead of covering the full range $(\phi_\text{FV},\phi_\text{TV})$, the bounce action will run from the false vacuum $\phi_\text{FV}$ up to a certain \emph{exit point} $\phi_0$, located between the true vacuum and the \emph{equivalence point} $\phi_\text{eq}$ --- where the potential has the same value as in the false vacuum.
Numerically, we determine $\phi_\text{eq}$ by solving $V(\phi_\text{eq})=V(\phi_\text{FV})$ with \texttt{FindRoot}, initialized at the midpoint $(\phi_\text{FV}+\phi_\text{TV})/2$.
This feature can aid the segmentation of the field space, since, once the number of segments used to construct the polygonal bounce is fixed, restricting the range in field space increases the density of segments, improving the overall estimate of the bounce solution.
The improvement is particularly useful in proximity of the \emph{inflection point}, where potential barrier is very low, $\phi_\text{eq}$ is close to the false vacuum, and the exit point $\phi_0$ is far from the true vacuum. 

\begin{figure}
    \centering
    \includegraphics[width=0.47\linewidth]{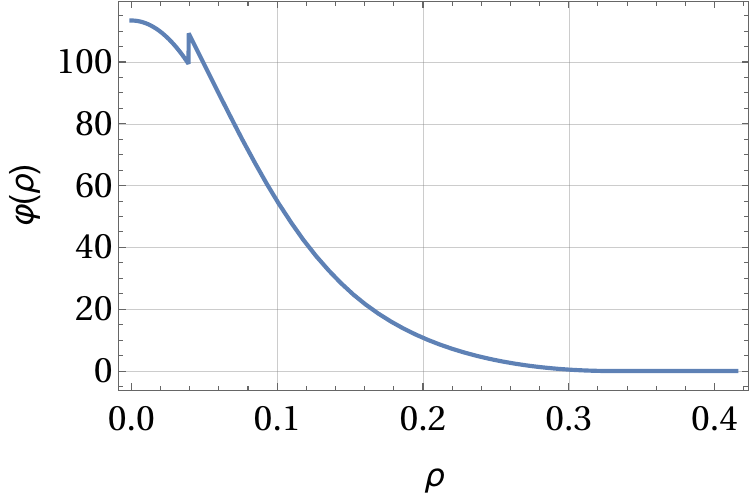}
    \hfill
    \includegraphics[width=0.47\linewidth]{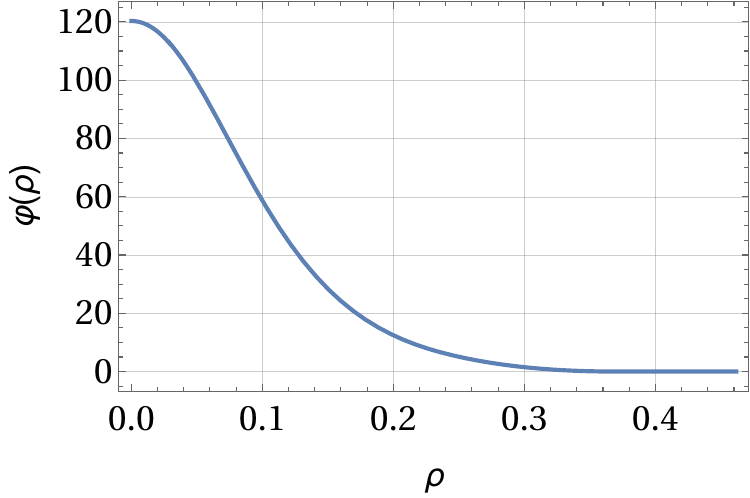}
    \caption{Improvement on the bounce profile by performing the shift of the exit point. 
    Here, $\rho$ is the Euclidean radius and $\phi(\rho)$ is the bubble profile. The bump on the left panel is an artifact of the polygonal bounce method implemented by \texttt{FindBounce} and is successfully eliminated in the right panel.}\label{fig:shift_bubble_profile}
\end{figure}

To estimate the exit point, we shift the second minimum by a quantity proportional to the distance between the true vacuum and the equivalence point:
\begin{equation}
    \phi_\text{TV} \rightarrow \phi_\text{TV}+s \, (\phi_\text{eq}-\phi_\text{TV}) \, ,
\end{equation}
where we defined the \emph{shifting parameter} $s$, which is set to 0.5 by default.
This procedure is implemented through the function \texttt{ShiftToExitPoint[T,V,Minima,Maximum]} and schematically illustrated in Fig.~\ref{fig:shift2ExitPoint}.
The available options are \texttt{"Shift"}$\rightarrow$0.5, which defines the shifting parameter $s$, and \texttt{"PrintShiftInfo"}$\rightarrow$\texttt{True}, which returns the shift.
In Fig.~\ref{fig:shift_bubble_profile} we report a visualization of the effect of the exit point shifting on the bounce profile.

\subsubsection{Action filter and refine} \label{subsec:action_filter_refine}
The \texttt{ActionFilter[\{$T,S/T$\}]} function is designed to filter out points from the list \texttt{\{$T,S/T$\}} that violate $\Delta^2(\log(S/T))/\Delta T^2<n \sigma$, where the maximum number of standard deviations allowed, $n$, is defined in the option \texttt{"stdMax"}. The default value is 1.

The function \texttt{ActionFilter[\{$T,S/T$\}]} takes as input the list \texttt{\{$T,S/T$\}} and constructs a new list of points \{$T,\Delta^2(\log(S/T))/\Delta T^2\}$, where $\Delta^2/\Delta T^2$ denotes a numerical second derivative. The standard deviation $\sigma$ of this dataset is then computed, and data points that violate $\Delta^2(\log(S/T))/\Delta T^2<n \sigma$ are filtered out, where the threshold parameter $n$, which defaults to 1, is specified by the option \texttt{"stdMax"}.

The \texttt{ActionRefine[V,\{$T,S/T$\},\{$\phi_1,\phi_2$\}]} function applies the filter function to the list \texttt{\{$T,S/T$\}}: it recomputes the action at the filtered-out temperatures from this list with the options passed to \texttt{ActionRefine}. By default, we improve on the computation of the action by lowering the \texttt{"ActionTolerance"} to $10^{-15}$. All the options of the \texttt{Action} function are available.

The \texttt{ActionRefineInflection[V,\{$\phi_1,\phi_2$\},\{$T_\text{miss},T_\text{hit}$\}]} function is designed for cases where the temperature region in proximity to the inflection point is relevant for the transition, for example if the nucleation temperature is there.
For such cases, it is possible that \texttt{FindBounce} fails at very low temperatures, for the linearly sampled set of temperatures.
The solution we provide is the implementation of a bisection method, by taking as initial extrema of the interval the highest temperature at which \texttt{FindBounce} fails, $T_\text{miss}$, and the lowest at which it succeeds, $T_\text{hit}$.
In such a way, we are able to push the temperature at which the action is evaluated as close as possible to the inflection point.
The available options are:
\begin{outline}
    \1 \texttt{"NHits"}$\rightarrow4$: the maximum number of newly successfully calculated pairs $\{T,S/T\}$, i.e. where the action has a numerical value;
    \1 \texttt{"MaxIterations"}$\rightarrow 20$: maximum number of iterations;
    \1 \texttt{"Keep"$\rightarrow$Last}: indices of the points to maintain from the list of newly obtained points. The default value \texttt{Last}, equivalent to \texttt{\{-1\}}, indicates we only include the last (lowest-temperature) point from the bisection algorithm.
\end{outline}
The reason to keep only the last datapoint is that this procedure generates an additional series of datapoints $\{T,S/T\}$ in a very narrow temperature range.
As a result, the temperature intervals in this zone are much smaller than the interval between $T_\text{miss}$ and $T_\text{hit}$.
This may cause an inaccurate cubic spline fit in the proximity of $T_\text{hit}$.

\subsubsection{Action fit}
The fitting procedure is performed by combining a Laurent polynomial in $(T_c-T)$ at high temperatures and a cubic spline at low and intermediate temperatures.
Continuity of the function, its first and second derivatives at the splitting point are imposed.
The user can pass the following options to the fit function:
\begin{outline}
    \1 \texttt{"Orders"}$\rightarrow$\texttt{Range[-2,1]}: determines the grade of the Laurent polynomial. The default option is motivated by analytical insights from the thin-wall approximation;\footnote{Close to the critical temperature, where the minima are almost degenerate, the Euclidean action at leading order in the expansion around $T_c$ is proportional to $(T_c-T)^{-2}$, see~\cref{eq:S3LO_Tc}.}
    \1 \texttt{Print}$\rightarrow$\texttt{False}: if \texttt{True}, returns the estimated variance of the fit.
\end{outline}
The two fitting modules above are wrapped into the \texttt{ActionFit[V,\{$\phi_1,\phi_2$\},\{T,S/T\},T$_c$]} module, which has the following options:
\begin{outline}
    \1 \texttt{"Data"}$\rightarrow$\texttt{None}: the list of data points \texttt{\{T,S/T\}};
    \1 \texttt{"NActionPoints"}$\rightarrow$31: if \texttt{"Data"} is \texttt{None}, this option gives the number of points at which the action is evaluated, uniformly distributed between the minimum and maximum temperature;
    \1 \texttt{"ActionMethod"}$\rightarrow$\texttt{"PWLaurent"}: selects the method to be used for the fit. The default value consists of a spline fit at low and intermediate temperatures, and a Laurent polynomial at high temperatures. Other possibilities are \texttt{Interpolation}, which is a spline fit only, or \texttt{"Laurent"}, which is a Laurent polynomial fit only;
    \1 \texttt{"StopAtFailure"}$\rightarrow$\texttt{False}: if the calculation of the Euclidean action via \texttt{FindBounce} fails, the module stops;
    \1 \texttt{"RefineInflection"}$\rightarrow$\texttt{\{True,"NHits"$\rightarrow$4,"MaxIterations"$\rightarrow$12,"Keep"$\rightarrow$Last\}}: the first argument indicates whether the \texttt{ActionRefineInflection} function is activated, implementing the bisection method to refine the computation of the action near the inflection point (see Sec.~\ref{subsec:action_filter_refine});
    \1 \texttt{"Refine"}$\rightarrow$\texttt{False}: determines whether the filter of the Euclidean action is applied (see Sec.~\ref{subsec:action_filter_refine});
    \1 \texttt{"PlotAction"}$\rightarrow$\texttt{False}: determines whether the plot of the Euclidean action is shown in the output;
    \1 \texttt{ProgressIndicator}$\rightarrow$\texttt{True}: determines whether to display a progress bar, showing the temperature at which the action is being evaluated.
\end{outline}
The function \texttt{PlotAction[S(T)/T,\{T$_\text{min}$T$_\text{max}$\}]} plots the Euclidean action as a function of temperature.
The first input is the action as a function of temperature, while the second input specifies the temperature range.
The available options are:
\begin{outline}
    \1 \texttt{"Data"}$\rightarrow$\texttt{\{\}}: specifies the list of values \{T,S/T\} to be visualized via a \texttt{ListPlot};
    \1 \texttt{"Temperatures"}$\rightarrow$\texttt{<||>}: specifies the relevant temperatures to be shown in the plot, in the form of an association with available keys \texttt{"Tc","Tn","Tp"};
    \1 \texttt{PlotLegends}$\rightarrow$\texttt{Automatic}: makes the legend for the action plot, including the \{T,S/T\} values, the fit, and the relevant temperatures.
\end{outline}

\subsubsection{Analytical action}\label{subsec:manual_action_analytic}
For polynomial potentials up to order four, an analytical form for the Euclidean action is available, see \ref{app:analytics}.
This can be used for the calculation of the nucleation and percolation temperatures, instead of the action fit.
For a suitable polynomial potential with numerical parameters, the function \texttt{ActionPolynomial[V]} returns the Euclidean action from \cref{eq:analytical_thermal_action}.
If the potential is not of the correct form, an error is returned.
This function is contained in the \texttt{Models} sub-package, to allow the user to easily customize it.

\subsection{Nucleation temperature}
The function \texttt{FindNucleation[$V,\{T_\text{min}, T_\text{max}\},\{\phi_{1},\phi_{2}\}$]} searches for a nucleation temperature in the given range and for the given phases.
Different criteria and methods are available, as listed below:
\begin{outline}
    \1 \texttt{"NucleationCriterion"}$\rightarrow$\texttt{"DecayOverHubble"}: selects the nucleation criterion. We provide the following criteria:
        \2 \texttt{"DecayOverHubble"} solves 
        \cref{eq:GammaH-4};
        \2 \texttt{"IntegralDecay"} uses the integral criterion \cref{eq:Tn_definition}. This criterion requires to pass the action function via \texttt{"ActionFunction"} method (see \texttt{"NucleationMethod"});
        \2 \texttt{\{"ActionValue"$\to$k\}} uses $S_3/T=k$, analogous to \cref{eq:Tn_condition_EW} for the EW case;
    \1 \texttt{"NucleationMethod"}$\rightarrow$\texttt{"Bisection"}: picks an algorithm to implement the nucleation criterion. It may correspond to one of the following:
        \2  \texttt{"Bisection"} implements the bisection algorithm in the given temperature range. Adding \texttt{"PrintIterations"$\to$True}, the result of each iteration is printed;
        \2  \texttt{FindRoot} uses the built-in root solver of \Math;
        \2  \texttt{\{"ActionFunction"$\to$actionFunction,"Tc"$\to T_c$,"TnEstimate"$\to T_n^\text{estimate}$\}} passes the action function $S_3(T)/T$, presumably constructed with \texttt{ActionFit}. This method is required for the \texttt{"IntegralDecay"} criterion;
    \1 \texttt{AccuracyGoal}$\rightarrow1$: specifies the number of effective digits of accuracy;
    \1 \texttt{PrecisionGoal}$\rightarrow\infty$: specifies the number of effective digits of precision;
    \1 \texttt{Return}$\rightarrow$\texttt{"TnRule"}: determines the output of the function. The default value returns the nucleation temperature as a substitution rule. Additional outputs, which can be passed as a list, are:
        \2 \texttt{"STnRule"}: returns the Euclidean action at the nucleation temperature as a substitution rule;
        \2 \texttt{"Tn"}: returns the numerical value of the nucleation temperature;
        \2 \texttt{"NucleationAction"}: returns the numerical value of Euclidean action at the nucleation temperature;
		\2 \texttt{"NucleationDecayOverHubble"}: returns the numerical value of $\Gamma/H^4$ at the nucleation temperature;
		\2 \texttt{"NucleationIntegralDecay"}: returns the numerical value of the integral $\int_{T_n}^{T_c} (\di T/T)\, (\Gamma/H^4)$;
    \1 \texttt{ProgressIndicator}$\rightarrow$\texttt{True}: determines whether to display a progress bar;
    \1 \texttt{Print}$\rightarrow$\texttt{False}: prints \texttt{"Computing Tn"} at the beginning of the execution;
\end{outline}

\subsubsection{Analytical nucleation temperature: CFF model}
For specific cases in which the potential is similar to the one for the CFF model (see~\cref{eq:param_T_scaling}), the function \texttt{NucleationCFF[V,k]} implements the analytical estimates for the nucleation temperature, described in \ref{subsec:Tn_analytic}.
Given the potential \texttt{V} and the target value for the action-over-temperature \texttt{k}, the algorithm extracts the polynomial coefficients and estimates the nucleation temperature in both the expansion around the critical temperature $T_c$, \cref{eq:Tn_from_Tc_generic}, and around the inflection point $T_0$, \cref{eq:Tn_from_T0_generic}.
With the default option \texttt{"Expansion"$\rightarrow$Automatic}, the best estimate is returned, based on the value of the expansion parameter given in \cref{eq:epsilon_alpha}. See discussion in \ref{subsec:Tn_analytic} for more details.
Alternatively, the option values \texttt{"Inflection"} and \texttt{"Critical"} can be used to select a specific expansion.
The estimate for $T_n$ obtained with this function 
can be passed to the \texttt{SearchPhases} module with the option \texttt{"TnEstimate"$\rightarrow T_n^\text{est}$} (see Section~\ref{subsec:SearchPhases}), reducing the computation time.

\subsection{Percolation temperature}
The function \texttt{FindPercolation[actionFunction,\{T$_\text{c}$,T$_\text{p,Guess}$\},$v_w$]} calculates the percolation temperature by evaluating the integral in~\cref{eq:I_percolation}, starting from the initial guess $T_\text{p,Guess}$.
Given that we want to numerically evaluate the integral, and the quantity we are looking for is the lower boundary of integration, in the definition of the integral function this temperature is specified to be a numerical value via the command \texttt{?NumericQ}.
The check for the successful completion of the transition given by the percolation condition~\cref{eq:perc_condition} is also included.
The options for the calculation of the percolation temperature are the following:
\begin{outline}
    \1 \texttt{Method}$\rightarrow$\texttt{FindRoot}: the method to solve for the percolation temperature. An initial guess for the percolation temperature $T_\text{p,Guess}$ is needed, which is provided by the calculated nucleation temperature. For numerical purposes, we solve~$I(T)=0.34$ from~\cref{eq:I_percolation} by taking the logarithm of both sides, given that the LHS is a rapidly changing function of temperature. Alternatively, with \texttt{"Bisection"}, a bisection algorithm can be used, where the temperature range has to be provided (see the option \texttt{"TRange"} below);
    \1 \texttt{"Target"}$\rightarrow 0.34$: the numerical value of the integral we are aiming for;
    \1 \texttt{"TRange"}$\rightarrow$\texttt{\{\}}: list consiting of lower and upper boundary of temperatures, needed for the bisection method;
    \1 \texttt{Return}$\rightarrow$\texttt{"Tp"}: selects the output format, which by default is the numerical value of $T_p$. It can include \texttt{"TpRule"} which returns a substitution rule, \texttt{"Value"} that returns the value of the percolation integral \cref{eq:I_percolation}, and \texttt{"Action"}, corresponding to the value of the Euclidean action at $T_p$;
    \1 \texttt{ProgressIndicator}$\rightarrow$\texttt{True}: determines whether to display progress bars.
\end{outline}

\subsection{SearchPhases}\label{subsec:SearchPhases}
\texttt{SearchPhases} is a wrapper module that, given the potential and a pair of phases, searches for phase transitions and computes the resulting GW spectra. It returns a \texttt{Transition} object, which stores all the quantities calculated in the process.
To compute the spectra, it calls the function \texttt{ComputeGW} defined in the \texttt{GW} sub-package (see Section~\ref{subsec:manual_GW}), by providing the required phase transition parameters. 
Since this module performs a sequence of computations, we verify the result type\footnote{
In \Math, this means ensuring that expressions have the correct \texttt{Head}.
}
at each step: if a result has the wrong type, the computation halts immediately, returning the current (partial) outcome.

The options specific to \texttt{SearchPhases} are the following:
\begin{outline}
    \1 \texttt{"TnEstimate"}$\rightarrow$\texttt{Automatic}: an estimate of the nucleation temperature, used to define the region where the action should be fitted. If set to \texttt{Automatic}, the nucleation temperature is estimated by a bisection method in the region $(T_\text{min},T_c)$, where $T_\text{min}$ is the minimum temperature at which the phases overlap;
    \1 \texttt{"ActionFunction"}$\rightarrow$\texttt{Automatic}: if the Euclidean action function $S_3(T)/T$ has already been computed, it can be provided with \texttt{"ActionFunction"->function}. For simple potentials, it might be possible to construct an analytical form of the action, as is the case for the CFF model discussed in \ref{app:analytics}. 
    By default, the function is applied at the steps labelled \texttt{"TnEstimate"}, \texttt{"Tn"} and \texttt{"Tp"} (see Fig.~\ref{fig:package_structure}), avoiding the use of \texttt{FindBounce} altogether. However, users can control where to apply the provided action function by including the argument \texttt{"ApplyTo"}, followed by a list including any combination of the above. For instance, to apply the action function only at the computation of the nucleation temperature and its estimate, use the following:
    \texttt{"ActionFunction"->\{function,"ApplyTo"->\{"TnEstimate","Tn"\}\}}
    ;
    \1 \texttt{"$\Delta$TFraction"}$\rightarrow$0.8: determines the range of temperatures used for the action fitting procedure. Given the critical temperature $T_c$ and the initial estimate of the nucleation temperature $T_n^{(0)}$, the upper and lower bounds are given by $T_{\substack{\text{max}\\\text{min}}}=T_n^{(0)}\pm \Delta \text{TFraction} (T_c-T_n^{(0)})$. With the default value, the interval is centred around $T_n^{(0)}$ and covers 80\% of the temperature range towards the critical temperature, and symmetrically for lower temperatures;
    \1 \texttt{"TransitionTemperature"}$\rightarrow$\texttt{"Tp"}: selects the temperature at which the transition parameters $\alpha$, $\beta/H$, \dots\ are computed. It should match either \texttt{"Tc"}, \texttt{"Tn"} or \texttt{"Tp"};
    \1 \texttt{"Metadata"}$\rightarrow$\texttt{\{\}}: any additional info the user may want to store in a \texttt{Transition} object. For example, to add the benchmark parameters: \texttt{"Metadata"->\{"mu"->100.,"lambda"->0.1\}};
	\1 \texttt{UpTo}$\rightarrow$\texttt{All}: stops the sequence of computations at the given argument, returning an \texttt{Association} of the results obtained so far. For example, to stop the algorithm at the computation of the nucleation temperature, use \texttt{UpTo->"Tn"}. The default value (\texttt{All}), performs all computations;
    \1 \texttt{"Plots"}$\rightarrow$\texttt{None}: if defined, determines which quantities among the phases $\phi(T)$, the action function $S_3(T)/T$ and the GW spectra $h^2\Omega_\mathrm{GW}(f)$ should be plotted. Can be set to \texttt{None},\footnote{
    During an evaluation, \emph{dynamic updating} features --- such as hovering over plots ---- may cause issues (including crashes). Therefore, the default setting for \texttt{"Plots"} is \texttt{None}.
    } \texttt{All}, or a list including a combination of \texttt{"Action"} and \texttt{"GW"}. For example, \texttt{"Plots"->\{"GW"\}} will only plot the GW spectra.
    If set to the empty value \texttt{\{\}}, the options of the wrapped functions determine which quantities should be plotted.
\end{outline}

\subsection{SearchPotential}
This function is the global wrapper that incorporates both the phase tracing procedure and the \texttt{SearchPhases} between each pair of phases. All the options of the subfunctions are automatically available, together with the following options:
\begin{outline}
    \1 \texttt{"PrintNoCritical"}$\rightarrow$\texttt{False}: if set to \texttt{True}, enables information messages when there is no overlap or critical temperature between a pair of phases;
    \1 \texttt{Dataset}$\rightarrow$\texttt{False}: if set to \texttt{True}, returns a \Math\ \texttt{Dataset}, where rows correspond to transitions, while columns correspond to transition parameters;
    \1 \texttt{"Plots"}$\rightarrow$\texttt{None}: determines which plots to include in the output. It works in the same way as the corresponding option in \texttt{SearchPhases}, with the added possibility to generate the phase diagram using \texttt{"PhaseDiagram"}.
\end{outline}

\subsection{Miscellaneous features} \label{subsec:manual_miscellaneous}
Here we describe additional features of \packageName, which do not directly involve any of the high-level functions.
\begin{outline}
    \1 The effective number of relativistic degrees of freedom, introduced in \cref{eq:Hubble} and typically denoted by $g_*$, is defined by the symbol \texttt{RelativisticDOF}, which defaults to the EW-scale Standard Model value of 106.75.
    Although this quantity depends on temperature \cite{Laine:2015kra,Husdal:2016haj}, $g_*$ is assumed to remain approximately constant during the phase transition and to be known at the transition temperature.
    \1 In order to simplify the numerics, and to minimize errors due to machine-number precision, the function \texttt{DefineUnits["unit"]} allows to set the energy unit, modifying the values of both the \texttt{\$Unit} symbol and the numerical value of the reduced Planck mass \texttt{\$PlanckMassN}.\footnote{
    These symbols have attribute \texttt{Protected}, meaning they cannot be modified directly.
    } While the default unit is $1~\text{GeV}$, to change it for example to $1~\text{keV}$ one must run
\begin{mmaCell}[leftmargin=3.1em,  labelsep=.5em, defined=DefineUnits]{Input}
DefineUnits["keV"]
\end{mmaCell}
    To obtain the exact string matching a certain unit, the user should first check the output of \texttt{Quantity["unit"]}.
    \1 \packageName\ functions print several information about their internal computations. While some of these messages can be turned off by passing the argument \texttt{Print->False}, the overall output can be disabled by setting \texttt{\$PT2GWPrint=False}.
    This does not affect the behavior of the \texttt{ProgressIndicator}s and plots.
    \1 \Math\ may display several error messages, which do not affect the result of computations within \packageName. \texttt{NewMessageGroup["groupName"]:>{messages}} allows to define groups of messages that can be switched off (on) with \texttt{Off["groupName"]} (\texttt{On["groupName"]}). A \texttt{"PT2GW"} message group is created and switched off by default.
    \1 Code autocompletion is enabled by \texttt{AutoComplete} (not to be mistaken for the built-in symbol \texttt{Autocomplete}). For example, \texttt{AutoComplete[function,\{f1,f2,\dots\}]} suggests to autocomplete \texttt{function} with the options for the given list of functions.
    \1 For the GW templates, the frequency unit is specified by \texttt{\$FrequencyUnit}, which defaults to \texttt{"Hz"}.
    The user can directly redefine \texttt{\$FrequencyUnit}, as well as $\varepsilon$ and $\mathcal{N}$ from the \texttt{GW} sub-package.
    \1 The function \texttt{PlotGW} plots the spectra as a function of frequency by taking as input an \texttt{Association} of GW spectral functions, where the keys should correspond to \texttt{"Collisions"}, \texttt{"Soundwaves"} and \texttt{"Turbulence"}. The user is expected to compute this \texttt{Association} by means of the \texttt{GW} sub-package (see Section~\ref{subsec:manual_GW} below). When multiple spectra are provided, the combined spectrum is also included. If not explicitly defined, the frequency range is determined automatically, such that the frequency corresponding to the peak amplitude is centred.
    The available options are:
        \2 \texttt{"FrequencyRange"}$\rightarrow$\texttt{Automatic}: sets the range on the horizontal axis;
        \2 \texttt{"h2OmegaRange"}$\rightarrow$\texttt{Automatic}: sets the range on the vertical axis;
        \2 \texttt{"Sources"}$\rightarrow$\texttt{\{"Collisions","Soundwaves","Turbulence"\}}: specifies the spectra of individual GW contributions that are shown;
        \2 \texttt{"Detectors"}$\rightarrow$\texttt{\{"LISA PISC","DECIGO PISC","BBO PISC"\}}: specifies the sensitivity curves of which detectors are shown. The plotting function \texttt{PlotGW} internally calls \texttt{PlotGWSensitivities} (see below);
        \2 \texttt{"GWPeak"}$\rightarrow$\texttt{True}: specifies whether the peak amplitude and frequency should be highlighted.
    \1 The function \texttt{PlotGWSensitivities[\{fmin,fmax\},detectors]} outputs the plot of the sensitivity curves for the experiments chosen by the user. The sensitivity curves are defined in the \texttt{GW} sub-package, and stored in the variable \texttt{GWSensitivities} (see Section~\ref{subsec:manual_GW} below). \texttt{GWSensitivities["List"]} gives the full list of available detector sensitivities.
\end{outline}

\subsection{The \texttt{GW} sub-package} \label{subsec:manual_GW}
The computations related to gravitational waves, i.e. the efficiency factors, the GW power spectra and the detector sensitivity curves, are placed in a separate \texttt{Mathematica} notebook, named \texttt{GW}. This sub-package is automatically loaded with \packageName, and its options can be passed to the functions \texttt{SearchPhases} and \texttt{SearchPotential} above.

The \texttt{ComputeGW[$T_*$,$\alpha$,$\beta/H$,$v_w$,$g_*$,unit]} function computes the power spectrum by taking the phase transition parameters $\alpha$ and $\beta/H$ at the transition temperature $T_*$, the wall velocity $v_w,$ and the number of relativistic degrees of freedom $ g_*$ as input. The \texttt{unit} specifies the energy unit. Independently of \packageName, the result of the computations are stored in the association \texttt{GWData}.
The available options are:
\begin{outline}
    \1 \texttt{"CollisionData"}$\rightarrow$\texttt{None}: collects the information about particles acquiring mass by crossing the bubble wall in the following format
\begin{mmaCell}[leftmargin=3.1em,  labelsep=.5em, defined={MassFunction,V,Tn,Tp,g}]{Input}
"CollisionData"->\{
    "GaugeCouplings"->\{g\},
    "MassFunctions"->\{"Bosons"->\{MassFunction[V]\}\},
    "T0"->Tn,"T"->Tp
    \}
\end{mmaCell}
    where \texttt{"T0"} and \texttt{"T"} are the temperatures corresponding to $T_0$ and $T_*$ in \cref{eq:Rstar_R0}, respectively. In this example, we set them to the numerical values for the nucleation and percolation temperature \texttt{Tn} and \texttt{Tp}. When passing this option to \texttt{SearchPotential} (or \texttt{SearchPhases}), one can use instead the labels \texttt{"Tn"} and \texttt{"Tp"}, which tell \packageName\ to substitute them for their numerical values, once computed.
    \texttt{"CollisionData"} is required to compute the efficiency factor for the GW production from bubble wall collisions, $\kappa_\text{col}$. Alternatively, the user can provide directly a numerical value:
\begin{mmaCell}[leftmargin=3.1em,  labelsep=.5em, defined={MassFunction}]{Input}
"CollisionData"->\{"κcol" -> 0.1\}
\end{mmaCell}     
     If $\kappa_\text{col}$ is not provided --- as per default option --- the only contributions to the GW spectrum are from sound waves and MHD turbulence in the plasma;
    \1 \texttt{"Sources"} $\rightarrow$ \texttt{\{"Collisions","Soundwaves","Turbulence","Combined"\}}: selects the sources to be included in the spectrum;
    \1 \texttt{"SpeedOfSound"}$\rightarrow 1/\sqrt{3}$: speed of sound in the plasma;
    \1 \texttt{"$\varepsilon$"}$\rightarrow0.5$: MHD efficiency parameter;
    \1 \texttt{"$\textit{N}$"}$\rightarrow 2$: number of eddy turnover times for turbulence;
    \1 \texttt{"ShapeParameters"}$\rightarrow$$\Omega_\text{pars}$: following the notation of Ref.~\cite{Caprini:2024hue}, it is a dataset containing the parameters $n_1$, $n_2$, $n_3$, $a_1$, $a_2$, which determine the functional dependence on the frequency of the spectra in~\cref{eq:Omega}.
\end{outline}
The function \texttt{GWSensitivities} stores several detector sensitivity curves, for current and future detectors. These are functions of the form $h^2\Omega_\text{GW}(f)$, where $f$ is the frequency in Hertz and $h^2\Omega_\text{GW}$ is the gravitational-wave energy density power spectrum.
We include PLISCs\footnote{
We constructed PLISCs by interpolating a subset of the data provided in Ref.~\cite{Schmitz:2020syl}.
} for
BBO, CE, DECIGO, EPTA, ET, HL (LIGO Hanford and LIGO Livingston), HLV (LIGO Hanford, LIGO Livingston and VIRGO), HLVK (LIGO Hanford, LIGO Livingston, VIRGO and KAGRA), HLVO2 (LIGO Hanford, LIGO Livingston and VIRGO, at a sensitivity representative of observing run 2), IPTA, LISA, NANOGrav, PPTA, SKA, 
and the semi-analytical PISCs for LISA, DECIGO and BBO from Ref.~\cite{Schmitz:2020syl}. For the latter, the GW source must be specified through the \texttt{"Source"} option. For example,
\begin{mmaCell}[leftmargin=3.1em,  labelsep=.5em, defined={GWSensitivities}]{Input}
GWSensitivities[f,"LISA PISC","Source"->"Soundwaves"]
\end{mmaCell}
The full list of detectors can be obtained with \texttt{GWSensitivities["List"]}. 
These sensitivities can be passed to \texttt{PlotGWSensitivities} and \texttt{SearchPhases/SearchPotential} through the \texttt{"Detectors"} option.
For example, the following line of code plots all available sensitivity curves:
\begin{mmaCell}[leftmargin=3.1em,  labelsep=.5em, defined=PlotGWSensitivities]{Input}
PlotGWSensitivities[10.^\{-9, 4\}, All]
\end{mmaCell}
\mmaCellGraphics[leftmargin=4.1em,labelsep=1em,ig={width=.85\textwidth}]{Output}{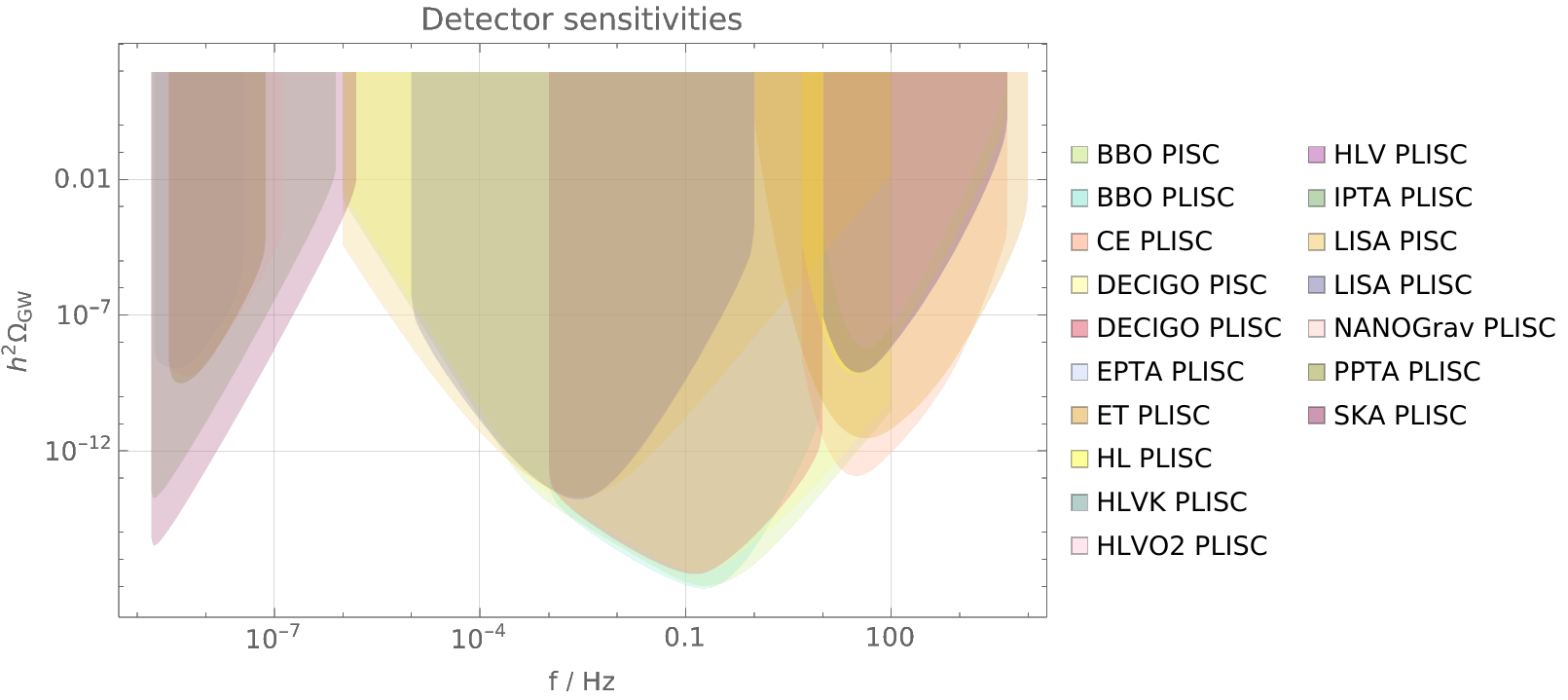}

Furthermore, the user can define their own detector sensitivity in the following way:
\begin{mmaCell}[leftmargin=3.1em,  labelsep=.5em, defined={GWSensitivities,OptionsPattern},pattern={f_,f}]{Input}
GWSensitivities[f_,"myDetector",OptionsPattern[]]:= 
    myFunction[f]
\end{mmaCell}

To customize the GW spectral templates, the user must redefine \texttt{ComputeGW}, ensuring that its output includes an \texttt{"h2Omega"} element. This element should be an \texttt{Association} of pure functions of frequency --- one per GW source --- matching the structure of the default template.

\subsection{The \texttt{DRTools} sub-package} \label{subsec:DRTools}
The \texttt{DRTools} utility collects the output of \texttt{DRalgo} \cite{Ekstedt:2022bff} into a closed-form thermal potential, to be used as input for \packageName. To load it, the user should run the following command:
\begin{mmaCell}[leftmargin=3.1em,  labelsep=.5em, defined={PT2GW,DRTools,m},pattern={f_,f}]{Input}
<<PT2GW/DRTools.m
\end{mmaCell}

After running \texttt{DRalgo} to obtain the effective masses, couplings, and parameters, the function \texttt{ComputeDRPotential[benchmark\_parameters, energy\_range]} computes the effective potential $V(\phi,T)$, including RG evolution.
As a minimal example, to obtain a potential for the Dark Abelian Higgs model of Section~\ref{subsec:ah_model}, the user should run
\begin{mmaCell}[leftmargin=3.1em,  labelsep=.5em, defined={ComputeDRPotential,SubRulesAppend},pattern={f_,f}]{Input}
(* values for g^2, lambda, mass^2 *)
benchmark=\{0.7^2,0.01,-4.^2\};
(* reference energy, min, max *)
range=\{40.,.1,1000.\};
(* substitution rules *)
subRules=SubRulesAppend[\{g->Sqrt[gsq]\}];
ComputeDRPotential[benchmark,range,
    "SubRules"->subRules
    ]
\end{mmaCell}
Here we defined the values for the model parameters $g^2$, $\lambda$, and $m^2=-\lambda v^2$ (where $v$ is the VEV of the dark scalar) at the reference energy scale in the \texttt{benchmark} variable, and the value for the reference scale and its range in the \texttt{range} variable.

The substitution rules must be defined for any coupling where the left-hand side of the $\beta$ function is not a symbol.
In the current example, \texttt{DRalgo} produces the $\beta$ function $g^2\rightarrow g^4/(24\pi^2)$, thus requiring the substitution $g\rightarrow \sqrt{gsq}$.
Since analogous rules must be applied to the 3D soft (and ultrasoft) parameters, the function \texttt{SubRulesAppend} is provided to automatize the process. 
The user should inspect the result of the \texttt{DRalgo} function \texttt{BetaFunction4D[]} to determine the appropriate rules.

The options available to \texttt{ComputeDRPotential} are:
\begin{outline}
    \1 \texttt{"SubRules"}$\rightarrow$\texttt{None}: if defined, sets the substitution rules required to generate $\beta$ functions and compute all quantities in the 3D theory. Alternatively, these can be loaded from a file;
    \1 \texttt{"LoadDRFrom"}$\rightarrow$\texttt{None}: allows to load from a file the expressions for the 3D theory;
    \1 \texttt{"OrderDR"}$\rightarrow$\texttt{"NLO"}: sets the order of the dimensional reduction;
    \1 \texttt{"OrderVeff"}$\rightarrow$\texttt{"NNLO"}: determines the expansion order of the effective potential;
    \1 \texttt{"NumericRules"}$\rightarrow$\texttt{\{\}}: rules to be applied to any auxiliary parameter which does not have a numerical value (and does not evolve with the RG equations). For example, in the Dark Abelian Higgs model, such a parameter is the hypercharge: \texttt{\{Y$\rightarrow$1.\}} converts the symbol \texttt{Y} to its numerical value before performing the RG evolution;
    \1 \texttt{"PlotRG"}$\rightarrow$\texttt{False}: determines whether to plot the results of RG evolution.
\end{outline}
\texttt{ComputeDRPotential} returns the closed-form, 3D effective potential, and stores the quantities computed in the process in an \texttt{Association} called \texttt{DRExpressions}. This includes $\beta$ functions, 3D \emph{soft} and \emph{ultrasoft} expressions, RG solutions and the effective potential.

While the function \texttt{ComputeDRPotential} is a high-level wrapper designed to automatize the process outlined above, the algorithm can be followed step by step by directly accessing the lower-level functions \texttt{StoreDRExpressions}, \texttt{RGSolve}, \texttt{DRStep} and \texttt{DRPotentialN}:
\begin{outline}[enumerate]
    \1 \texttt{StoreDRExpressions} stores quantities computed with \texttt{DRalgo} in the association \texttt{DRExpressions}.
    The options for this functions are:
        \begin{itemize}
            \item \texttt{"SubRules"}$\rightarrow$\texttt{\{\}}: before storing the expressions, it applies the substitution rules to all of these, as explained at the beginning of this section;
            \item \texttt{"File"}$\rightarrow$\texttt{None}: if provided with a file path, it saves the expressions to the file, in the format of an \texttt{Association};
            \item \texttt{"US"}$\rightarrow$\texttt{True}: allows to store quantities at the \emph{ultrasoft} scale, if these were computed with \texttt{DRalgo} (for more details, please refer to Ref.~\cite{Ekstedt:2022bff}). It is still possible to store these quantities, while computing the effective potential at the \emph{soft} scale;
            \item \texttt{"PrintDR"}$\rightarrow$\texttt{False}: allows to display all the stored quantities in a neatly formatted table.
        \end{itemize} 
        This is the only step requiring the preliminary run of \texttt{DRalgo}. Later, the expressions can be loaded from \texttt{file.m} with \texttt{LoadDRExpressions[file.m]}.
    \1 \texttt{RGSolve[RGequations,benchmark,\{$\mu$,$\mu_\text{min}$,$\mu_\text{max}$\}]} solves numerically the RG equations by taking as input the list of equations, the boundary conditions and a list containing the reference, minimum and maximum for the renormalization scale. In practice, \texttt{RGSolve} is a wrapper of the built-in \texttt{NDSolve} function of \Math.
    Instead of the RG equations, one can also pass the $\beta$ functions defined in \texttt{DRExpressions} (with key \texttt{"BetaFunctions"}): the function \texttt{BetaToEquations} will automatically convert them to RG equations. For example, the expression $g\rightarrow \beta[g]$ is converted to $\mu g'[\mu]==\beta[g]$.
    The available options are:
    \begin{itemize}
        \item \texttt{"NumericRules"}$\rightarrow$\texttt{\{\}}: rules to be applied to any auxiliary parameter which does not have a numerical value. A non-numerical parameter will prevent the resolution of the RG equations;
        \item \texttt{Return}$\rightarrow$\texttt{Rule}: determines whether to return a list of rules or a list of expressions.
    \end{itemize}
    \1 \texttt{DRStep[$T$,s,parameters4D]} returns the numerical values for the masses, couplings and matching scales in the 3D effective theory at the temperature $T$. The constant \texttt{s} is a factor relating the temperature to the energy scale $\mu\equiv s \pi T$. The \texttt{parameters4D} are numerical rules for the 4D parameters, evolved to the scale $\mu$.
    The options are:
    \begin{itemize}
        \item \texttt{"OrderDR"}$\rightarrow$\texttt{"NLO"}: sets the order at which the dimensional reduction procedure is performed. Following the convention of \texttt{DRalgo}, this can be either \texttt{"LO"} or \texttt{"NLO"}, corresponding to the leading- and next-to-leading orders;
        \item \texttt{"US"}$\rightarrow$\texttt{True}: determines whether the dimensional reduction procedure is pushed to the ultrasoft scale;
        \item \texttt{"CheckDebyeMassSq"}$\rightarrow$\texttt{True}: enables to check for negative squared Debye masses, printing a warning message in case any is found;
        \item \texttt{"NumericRules"}$\rightarrow$\texttt{\{\}}: rules to be applied to any auxiliary parameter which does not have a numerical value.
    \end{itemize}
    \1 \texttt{DRPotentialN[$\phi$,$T$,RGsolutions,s]} gives the total potential with numerical values for the effective parameters, as a function of the scalar field $\phi$ and the temperature $T$. The solutions to the RG equations and the factor \texttt{s} must be provided. This function calls internally \texttt{DRStep}, which performs the dimensional reduction, and \texttt{DefineDRPotential}, which extracts the various orders of the analytic, effective potential, and stores them in the \texttt{DRExpressions} with key \texttt{"VTot"}. The options are:
    \begin{itemize}
        \item \texttt{"OrderVeff"}$\rightarrow$\texttt{"NNLO"}: sets the order at which the effective potential is returned. Similarly to \texttt{"OrderDR"}, this can be set to \texttt{"LO"}, \texttt{"NLO"} or \texttt{"NNLO"};
        \item \texttt{"RescaleTo4D"}$\rightarrow$\texttt{True}: determines whether to rescale the 3D effective theory to 4D according to \cref{eq:rescale_to_4D};
        \item \texttt{"NumericRules"}$\rightarrow$\texttt{\{\}}: rules to be applied to any auxiliary parameter which does not have a numerical value.
    \end{itemize}
\end{outline}
We provide three plotting functions to easily visualize the results from these computations:
\begin{outline}
    \1 \texttt{PlotRG[parameters,range]} plots the RG evolutions for a list of model parameters, in the given energy range. One can pass \texttt{All} as the \texttt{parameters} argument for a combined preview of all RG solutions;
    \1 \texttt{PlotDR[parameters,range,s]} plots the effective parameters in the given temperature range. The factor \texttt{s} must be provided;
    \1 \texttt{PlotHighTRatio[masses,range,s]} plots the maximum of the mass-to-energy scale ratio $m_\text{eff}/\mu$ in the given temperature range. This parameter can be useful to assess the perturbativity of the theory in the high-temperature regime.
\end{outline}
The option \texttt{"EnergyScaleSymbol"} allows the user to set the label of the horizontal axis, corresponding to the renormalization scale.

\section{Summary and Conclusions} \label{sec:conclusion}
\noindent
Gravitational wave astronomy is a well-established and highly promising field for studying the early Universe.
Significant experimental efforts have been devoted to the development of current and near-future GW interferometers, which have the potential to discover a SGWB.
Among the leading cosmological sources of a SGWB are cosmological FOPTs, predicted by many extensions of the Standard Model.

With this paper we launch \packageName, the first \Math\ package designed for the complete analysis of a cosmological FOPT, including the respective GW spectra.
Given the effective potential of a single-VEV model and the bubble wall velocity as inputs, \packageName\ performs the phase tracing,
computes the bounce solution and action, determines the relevant temperatures and phase transition parameters, and outputs the resulting GW power spectrum.
In order to illustrate the success of \packageName\ in the treatment of phenomenologically relevant models, we have presented results for the CFF model in \cref{subsec:cff_model} and for the Dark Abelian Higgs model in \cref{subsec:ah_model}.
This paper serves also as a manual for the user, with detailed descriptions of all the modules and functions implemented.
The package is designed for accessibility and ease of use, enabling the user to efficiently identify and characterize a model's phase transitions, and perform detailed analyses with minimal effort.
Any thermal scalar potential that features a FOPT, where a single field acquires a VEV, is suitable for \packageName.
We provide an example notebook that demonstrates how to implement a one-loop thermal potential in the daisy resummation approach.
Moreover, the use of dimensionally reduced potentials constructed with \texttt{DRalgo} is straightforward with our \texttt{DRTools} helper module.

\packageName\ features a modular structure designed to easily accommodate user-defined methods and integrate future packages and algorithms.
Following the release of \packageName-1.0.0, we plan to extend the package to include the treatment of a decay in multiple field directions.
%

\section*{Acknowledgments}
\noindent
The work of VB is supported by the United States Department of Energy Grant No. DE-SC0025477.
MM is supported by the Slovenian Research Agency's young researcher program under grant No. PR-11241.
MN is supported by the Slovenian Research Agency under the research core funding 
No.~P1-0035 and in part by the research grants N1-0253, J1-4389 and J1-60026.
MF and APM are supported by LIP and by the
Portuguese Foundation for Science and Technology (FCT), reference LA/P/0016/202 and from ERC-PT A-Projects ``Unveiling'', financed by PRR, NextGenerationEU.
MF is also directly funded by FCT through the doctoral program grant with the reference PRT/BD/154730/2023 (\url{https://doi.org/10.54499/PRT/BD/154730/2023}), within the scope of the ECIU University. MF acknowledges support from COST Action CA21106 through a Short-Term Scientific Mission to the Jo\v{z}ef Stefan Institute.
APM is also supported by FCT through the project with reference 2024.05617.CERN (\url{https://doi.org/10.54499/2024.05617.CERN}).

\appendix 
\section{Analytical bounce action and nucleation temperature} \label{app:analytics}
\noindent
In this Appendix we first briefly review the derivation of the Euclidean action for a single self-interacting scalar field using the thin-wall approximation, following Ref.~\cite{Matteini:2024xvg}, and then apply this result to calculate analytically the nucleation temperature.
\subsection{Thin-wall approximation}
The model features a quartic potential
\begin{align} \label{eq:generic_potential}
    V(\phi)=\frac{1}{2}m^2\phi^2+\eta\phi^3+\frac{1}{8}\lambda_C\phi^4 \, ,
\end{align}
which exhibits a FOPT.
The field and parameters have the following mass dimension, in $D$-dimensional spacetime
\begin{align} \label{eq:param_scaling}
    [\phi]=\frac{D-2}{2}\, , \quad [m^2]=2 \, , \quad [\eta]=\frac{6-D}{2}\, , \quad[\lambda_C]=4-D\, ,  \,
\end{align}
so that $[V]=D$.
This three-parameter theory can be recast into a single-parameter, dimensionless theory in the following way. 
First, introduce a dimensionless field and coupling
\begin{align}
    &\varphi\equiv\frac{2\eta}{m^2}\phi \, , &\epsal\equiv 1-\frac{\lambda_C m^2}{4\eta^2} \, .
\end{align}
Notice that both these quantities are dimensionless in any spacetime dimension $D$.
Then define a dimensionless potential $\tilde{V}$ as
\begin{align}
    \tilde{V}\equiv\frac{4\eta^2}{m^{6-D}}V \, ,
\end{align}
where
\begin{align}
    \tilde{V}(\varphi) = \frac{\varphi^2}{2}+\frac{\varphi^3}{2}+\frac{(1-\epsal)\varphi^4}{8} \, .
\end{align}
The reason behind the convention for the numerical factors in $\tilde{V}$ is that this matches exactly the potential studied in Ref.~\cite{Matteini:2024xvg}.
Notice that the action can then be written as 
\begin{align}
     S &= \Omega \frac{m^{6-D}}{4\eta^2} \tilde{S}_C(\epsal) \, ,
\end{align}
where $\Omega=2\pi^{D/2}/\Gamma(D/2)$ is the solid angle in $D$ dimensions and
$\tilde{S}_C(\epsal)$ is only a function of $\epsal$ via $\tilde{V}(\varphi)$.

The phase structure of the dimensionless theory is straightforward to obtain. For $\epsal< 1$, the true and false vacua are located at
\begin{align}
    &\phi_\TV = \frac{3+\sqrt{1+8\epsal}}{2(-1+\epsal)} \, , &\phi_\FV = 0 \, , 
\end{align}
and the top of the barrier is at
\begin{align}
    \phi_{\max} = \frac{3-\sqrt{1+8\epsal}}{2(-1+\epsal)} \, .
\end{align}
The two minima are degenerate for $\epsal=0$, while the potential becomes unbounded from below for $\epsal=1$, which corresponds to a vanishing quartic coupling $\lambda_C$.
We restrict the parameter space to $0\le\epsal<1$, where the false vacuum is at the origin and the transition proceeds to a vacuum away from the origin.

We perform the thin-wall expansion, i.e. an expansion in powers of $\epsal$ around $\epsal=0$, to obtain analytical results for the bounce profile and action, as explained in detail in Ref.~\cite{Matteini:2024xvg}.
There, it was shown that truncating the $\epsal$ expansion at second order gave a very good approximation for the Euclidean action over the whole range of parameter space, even away from the strict thin-wall regime. The result is
\begin{align}
\label{eq:ActionCubic2ndorder}
     \tilde{S}_C^{(2)}(\epsal) =\left( \frac{D-1}{3\epsal} \right)^{D-1} \frac{2}{3D}\left(1+ \frac{3D+8}{2}\epsal+ \frac{9 D^3 -11D^2 +(138 -12\pi^2)D -64}{8 (D-1)} \epsal^2\right) \, .
\end{align}
The full Euclidean action, including the prefactor, is then given by
\begin{align}
S & = \frac{32\pi}{81} \frac{m^3}{4\eta^2} \frac{1}{\epsal^2} 
\left[1 + \frac{17}{2} \epsal + \left(\frac{247}{8} - \frac{9\pi^2}{4} \right) \epsal^2 \right]  &  \text{for }D = 3 \, , \label{SsumD3} \\
S & = \frac{\pi^2}{3} \frac{m^2}{4\eta^2} \frac{1}{\epsal^3}  
\left[1 + 10 \epsal + \left(37 - 2\pi^2 \right) \epsal^2 \right] & \text{for }D = 4 \, .
\label{SsumD4}
\end{align}

\subsection{Applications to thermal theories}
In cosmological settings, the parameters of the potential will depend on the temperature: $\lambda_C\to c_4(T) \, , \eta\to c_3(T) \, , m^2\to c_2(T)$.
The Euclidean action in 3 dimensions, relevant for phase transitions at finite temperature, can be explicitly expressed as a function of the temperature 
\begin{align} \label{eq:analytical_thermal_action}
    S_3(T)=\frac{8 \pi  \, c_2(T)^{3/2}
   \left(\left(\frac{247}{8}-\frac{9 \pi ^2}{4}\right)
   \left(1-\frac{c_2(T)c_4(T)}{4 \, c_3(T)^2}\right)^2-\frac{17 \,
   c_2(T)c_4(T)}{8 \,
   c_3(T)^2}+\frac{19}{2}\right)}{81 \,
   c_3(T)^2
   \left(1-\frac{c_2(T)c_4(T)}{4 \, c_3(T)^2}\right)^2} \, .
\end{align}
The critical and nucleation temperatures of the phase transition can be studied by assuming the following temperature scaling
\begin{align} \label{eq:param_T_scaling}
    c_2(T) = a_2(T^2-T_0^2) \, , \quad c_3(T)^2=a_3^2(T+T_1)^2 \, ,  \quad c_4(T) = a_4 \, ,
\end{align}
derived by knowing the dimension of the parameters from~\cref{eq:param_scaling}.
An example of that is the CFF model presented in the main body.
Note that $a_4$ might still contain a $\log T $ dependence. Here, we ignore it in order to solve for the critical or nucleation temperatures analytically.
The notation highlights that both $T_0$ and $T_1$ have dimension of temperature or mass, with $T_0$ being related to the zero-temperature mass of the scalar.
The critical temperature is defined by
\begin{align}\label{eq:epsilon_alpha}
    \epsal=1-\frac{c_4(T)c_2(T)}{4c_3(T)^2} = 0 \, ,
\end{align}
which can be solved for $T$, giving
\begin{align} \label{eq:dimensionless_Tcrit}
    T_c = \frac{\sqrt{a_2^2 a_4^2 T_0^2+4 a_2 a_3^2 a_4 (T_1 ^2-T_0^2)}+4 a_3^2 T_1
   }{a_2 a_4-4 a_3^2} \, .
\end{align}
Moving away from the critical point, one can approach two different limits as $\epsal\to 1$, one driven by $c_4$ and one driven by $c_2$.
When $c_4\to 0$, the quartic term in the potential is vanishing, and one obtains an unbounded potential, which still admits a tunneling solution.
When $c_2\to 0$, one has $T^2\to T_0^2$ in a thermal theory and the potential develops an inflection point, where the false vacuum and the tip of the barrier merge.
The latter is the limit we will consider in the following.\footnote{In principle, one can consider an unbounded potential in a cosmological setting, assuming that it is stabilized at high energies by new physics. The bounce solution exists, and the Euclidean action is finite. For more details, please refer to Ref.~\cite{Matteini:2024xvg}.}

\subsection{Analytical calculation of the nucleation temperature}\label{subsec:Tn_analytic}
It is possible to obtain an analytical, benchmark-independent, formula for the nucleation temperature by expanding the LHS of the approximate criterion
\begin{equation}
    \frac{S_3}{T}=k
\end{equation}
in the appropriate temperature regime, where $k\simeq 140$ for transitions around the EW scale.
The two regimes are close to the critical temperature ($\epsal\to0$), and close to the inflection point temperature ($\epsal\to1$).
Once the nucleation temperature is calculated, one should check for self-consistency of the expansion a posteriori, using $\epsal(T_n)\simeq 0 $ for the expansion around the critical temperature and $\epsal(T_n)\simeq 1 $ for the expansion around the inflection point.
\subsubsection{Expansion around the critical temperature}
In order to expand close to the critical temperature, it is necessary to express the Euclidean action as a function of the critical temperature explicitly, using~\cref{eq:dimensionless_Tcrit}.
The higher orders in the expansion quickly increase the number of terms.
For this reason, here we only report the analytical result for the nucleation temperature by truncating~\cref{eq:ActionCubic2ndorder} at leading order in $\epsal$.\footnote{In \packageName, we include terms up to NNLO in the expansion.}
The result of the expansion for the Euclidean action is 
\begin{align} \label{eq:S3LO_Tc}
\frac{S_3}{T} & =\frac{256 \pi   a_3^2 T_c (T_1 +T_c) \left(\frac{a_3^2 }{ a_4}\right)^{3/2}}{81   \left(4 a_3^2-a_2 a_4\right)^2(T-T_c)^2} \, .
\end{align}
The \textit{analytical} result for the nucleation temperature is then
\begin{equation}\label{eq:Tn_from_Tc_generic}
    T_n = T_c-\frac{16 \sqrt{\pi } \sqrt{ a_4  \left(a_2 a_4-4 a_3^2\right)^2
   \left(\frac{a_3^2 }{ a_4}\right)^{5/2}T_c(T_1 +T_c)}}{9   \left(a_2 a_4-4
   a_3^2\right)^2 k^{1/2}} \, .
\end{equation}
For transitions at the EW scale, the nucleation temperature is given by 
\begin{equation}
    T_n=T_c-\frac{8 \sqrt{\frac{\pi }{35}} \sqrt{ a_4  \left(a_2 a_4-4 a_3^2\right)^2 \left(\frac{a_3^2 }{a_4}\right)^{5/2}T_c(T_1 +T_c)}}{9 
   \left(a_2 a_4-4 a_3^2\right)^2} \, .
\end{equation}

\subsubsection{Expansion around the inflection point temperature}
Around the inflection point, at $T=T_0$, up to NNLO in the thin-wall expansion, the Euclidean action is given by
\begin{align}
\frac{S_3}{T}=\frac{2 \sqrt{2} \pi  \left(323-18 \pi ^2\right) a_2^{3/2} (T_0 (T-T_0))^{3/2}}{81 a_3^2 T_0 (T_1 +T_0)^2} \, .
\end{align}
Notice that there is no dependence on the quartic coupling at this order in the expansion.
This might seem strange at first, but the reason lies in the fact that for tunneling close to the inflection point, the exit point is very close to the FV and to the top of the barrier, and far away from the TV, where the quartic term would be important.
We are again able to solve for the nucleation temperature \textit{analytically}
\begin{align} \label{eq:Tn_from_T0_generic}
    T_n = T_0+\frac{9\ 3^{2/3} \sqrt[3]{a_2^6 a_3^4 (T_1 +T_0)^4}}{2 \left(323 \pi -18 \pi ^3\right)^{2/3} a_2^3 \sqrt[3]{T_0}}  k^{2/3} \, .
\end{align}
For transitions at the EW scale, this reduces to
\begin{align}
    T_n =T_0 + \frac{9 \ \sqrt[3]{2} \ 105^{2/3} \sqrt[3]{a_2^6 a_3^4 (T_1 +T_0)^4}}{\left(323 \pi -18 \pi ^3\right)^{2/3} a_2^3 \sqrt[3]{T_0}} \, .
\end{align}

\subsection{Coupled fluid-scalar field model}
\begin{center}
    \begin{figure}[htb!]
        \centering
        \includegraphics[width=0.59\textwidth]{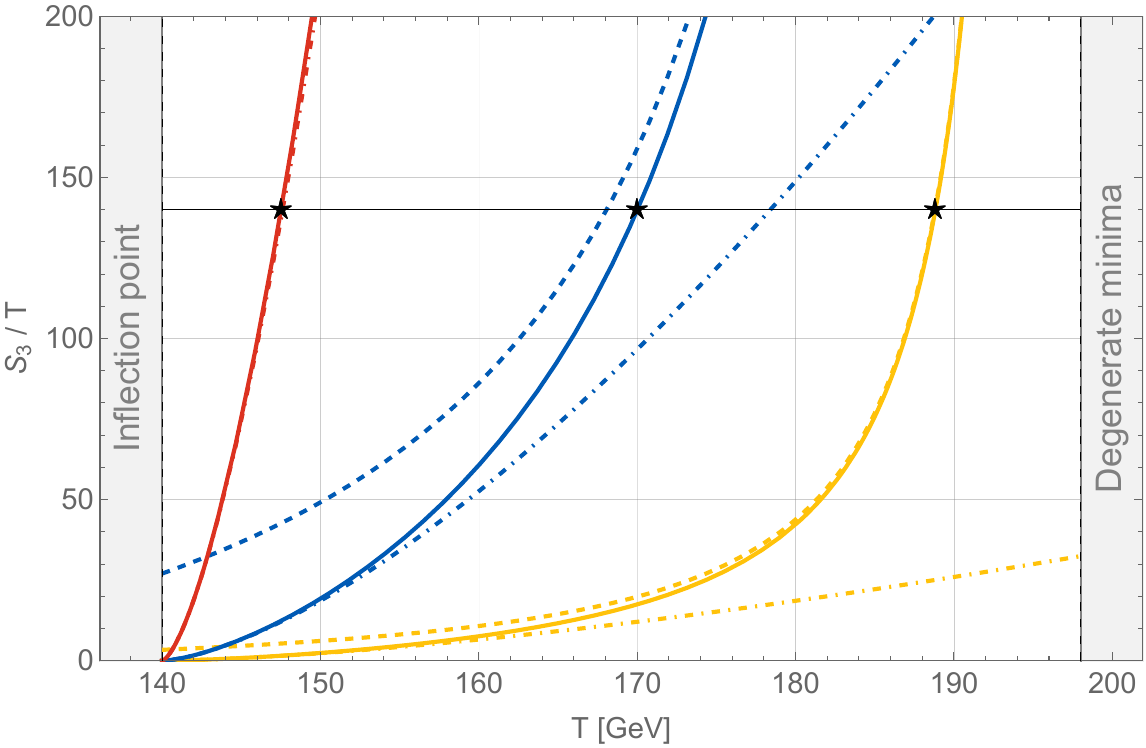}
        \caption{We show the action calculated analytically (solid line), via expansion around $T_c$ (dashed) and around $T_0$ (dot-dashed) for the CFF model. Benchmark points 1 (blue), 2 (yellow) and 3 (red) are shown. The black stars indicate the nucleation temperature obtained by intersecting the unexpanded analytical action with the horizontal line at $S_3/T=140$.}
        \label{fig:CFF_Tn_analytic}
    \end{figure}
\end{center}

We now apply the above expansions to the CFF model, and assess their reliability.
By comparing the temperature scaling of the coefficients~\cref{eq:param_T_scaling} and the CFF potential~\cref{eq:VCFF}, we obtain
\begin{align}
 a_2 = \gamma \, , \quad a_3 = -A/3 \, ,  \quad  a_4 = 2\lambda \, ,  \quad  T_1 = 0 \, . 
\end{align}
In Table~\ref{tab:Tn_estimates}, we compare the analytical results for the nucleation temperature from the expansion of the action around $T_c$ and $T_0$, with the nucleation temperature obtained by using the unexpanded analytical action and solving $S_3/T=140$.
To assess the reliability of the two expansions, we use the self-consistency relations: $\epsal(T_n)\simeq 0 $ for the expansion around the critical temperature, and $\epsal(T_n)\simeq 1 $ for the expansion around the inflection point.
We notice that, for benchmark point 1 where the nucleation temperature is in an intermediate regime, neither expansion is appropriate: $1-\epsal(T_n^{(T_c)})=0.61$ and $\epsal(T_n^{(T_0)})=0.23$.
For benchmark points 2 and 3, where the nucleation temperature is close to critical and to inflection, respectively, the appropriate expansion gives a reliable result.
This is confirmed by the self-consistency relations, which read $1-\epsal(T_n^{(T_c)})=0.90$ for benchmark point 2, and $\epsal(T_n^{(T_0)})=0.80$ for benchmark point 3.
\begin{table}[ht]
\centering
\begin{tabular}{ |c|c|c|c|c|c| } 
 \hline
 Benchmark 
 &  \, $T_0$ \, &  \, $T_c$  \,  &  \, $T_n^{(S/T)}$ \,  &  \, $T_n^{(T_c)}$  \,  & \,  $T_n^{(T_0)}$ \,  \\ 
 \hline
 1 
 & 140 & 197.99 & 170.02 &  168.08 & 178.41 \\ 
 2 
 & 140 & 197.99 & 188.83 & 188.80 & 293.65 \\
 3 
 & 140 & 197.99 & 147.52 & 129.93 & 147.62 \\
 \hline
\end{tabular}
\caption{Analytical estimates of $T_n$ from the expansions around $T_c$ and around $T_0$. All the temperatures are in GeV. 
}
\label{tab:Tn_estimates}
\end{table}
In Fig.~\ref{fig:CFF_Tn_analytic} we show the results of expansions of the action for the three benchmark points.
For cases when the expansion performs well, the result for the nucleation temperature from that expansion is reliable.

\bibliography{bib}

\end{document}